\documentclass[12pt]{article}

\usepackage{amssymb,amsmath,amsthm,amscd,latexsym}
\usepackage{amsfonts}
\usepackage{tikz}
\usetikzlibrary{arrows,positioning,shapes,fit,calc}
\usepackage{tikz-cd}
\usepackage[mathscr]{eucal}
\usepackage{color,hyperref}
\usepackage{graphicx}
\usepackage{longtable}
\usepackage{multirow}
\usepackage[ruled,vlined]{algorithm2e}
 \newtheorem{theorem}{Theorem}[section]
 \newtheorem{corollary}[theorem]{Corollary}
 \newtheorem{lemma}[theorem]{Lemma}
 
 \theoremstyle{definition}
 
 \theoremstyle{remark}

 \numberwithin{equation}{section}
\begin{document}

\title{New Singly and Doubly Even Binary $[72,36,12]$ Self-Dual Codes from   $M_2(R)G$ - Group Matrix Rings}
\author{ Adrian Korban \\
Department of Mathematical and Physical Sciences \\
University of Chester\\
Thornton Science Park, Pool Ln, Chester CH2 4NU, England \\
Serap \c{S}ahinkaya \\
Tarsus University, Faculty of Engineering \\ Department of Natural and Mathematical Sciences \\
Mersin, Turkey \\
Deniz Ustun \\
Tarsus University, Faculty of Engineering \\ Department of Computer Engineering \\
Mersin, Turkey}
 \maketitle

\begin{abstract}
In this work, we present a number of generator matrices of the form $[I_{2n} \ | \ \tau_k(v)],$ where $I_{kn}$ is the $kn \times kn$ identity matrix, $v$ is an element in the group matrix ring $M_2(R)G$ and where $R$ is a finite commutative Frobenius ring and $G$ is a finite group of order 18. We employ these generator matrices and search for binary $[72,36,12]$ self-dual codes directly over the finite field $\mathbb{F}_2.$ As a result, we find 134 Type I and 1 Type II codes of this length, with parameters in their weight enumerators that were not known in the literature before. We tabulate all of our findings.
\end{abstract}

\section{Introduction}\label{intro}

A search for new binary self-dual codes of different lengths is still an ongoing research area in algebraic coding theory. Many researchers have employed various techniques to search for binary self-dual codes of different lengths with new parameters in their weight enumerators. A classical technique is to consider a generator matrix of the form $[I_n \ | \ A_n],$ where $I_n$ is the $n \times n$ identity matrix and $A_n$ is some $n \times n$ matrix with entries from a finite field $\mathbb{F}_2.$ Of course, if we were to define the matrix $A_n$ in terms of $n^2$ independent variables, then only for the finite field $\mathbb{F}_2$ we would have a search field of $2^{n^2}$ which is not practical. Therefore, many researchers have considered matrices $A_n$ that are fully defined by the elements appearing in the first row - this reduces the search field from $2^{n^2}$ to $2^n.$ For example, one can consider the matrix $A_n$ to be a circulant or a reverse circulant matrix.

In \cite{Hurley1}, T. Hurley introduced a map $\sigma(v),$ where $v$ is an element in the group ring $RG$ with $|G|=n,$ that sends $v$ to an $n \times n$ matrix that is fully defined by the elements appearing in the first row - these elements are from the ring $R$ or finite field $\mathbb{F}_q.$ By employing different groups $G$ one can obtain different $n \times n$ matrices as images under the map $\sigma$ and this is the advantage of this map. In \cite{Gildea1}, the authors consider a generator matrix of the form $[I_n \ | \ \sigma(v)]$ for various groups $G$ to search for new binary self-dual codes of length 68 with a success. Recently in \cite{Dougherty1}, the authors extended the map $\sigma(v)$ so that $v \in RG$ gets sent to more complex $n \times n$ matrices that are also fully defined by the elements appearing in the first row. They name this map as $\Omega(v)$ and call the corresponding $n \times n$ matrices, the composite matrices - please see \cite{Dougherty1} for details. In \cite{Dougherty2}, generator matrices of the form $[I_n \ | \ \Omega(v)]$ are considered to search for binary self-dual codes.

The above two maps, $\sigma$ and $\Omega,$ both send an element $v$ from the group ring $RG$ to an $n \times n$ matrices that are fully defined by the elements appearing in the first rows. Recently in \cite{Dougherty3}, the authors extended the map $\sigma$ and considered elements from the group matrix ring $M_k(R)G$ rather than elements from the group ring $RG.$ They defined a map that sends an element from the group matrix ring $M_k(R)G$ to a $kn \times kn$ matrix over the ring $R.$ They called this map $\tau_k(v)$ - please see \cite{Dougherty3} for details. The advantage of this map is that it does not only depend on the choice of the group $G,$ but it also depends on the form of the elements from the matrix ring $M_k(R),$ that is, the form of the $k \times k$ matrices over $R.$ In this work, we employ the map $\tau_k(v)$ and consider generator matrices of the form $[I_{kn} \ | \tau_k(v)]$ with $k=2$ and groups of order 18 to search for binary self-dual codes with parameters $[72,36,12].$ We find many such codes with weight enumerators that were not known in the literature before.

The rest of the work is organized as follows. In Section~2, we give preliminary definitions and results on self-dual codes, special matrices, group rings and we also recall the the map $\tau_k(v)$ that was defined in \cite{Dougherty3}. In Section~3, we present a number of generator matrices of the form $[I_{kn} \ | \ \tau_k(v)]$ for $k=2$ and groups of order 18. For each generator matrix, we fix the $2 \times 2$ matrices by letting them be some special matrices that we define in Section~2. In Section~4, we employ the generator matrices from Section~3 and search for binary self-dual codes with parameters $[72,36,12].$ As a result we find 134 Type I and 1 Type II binary $[72,36,12]$ self-dual codes with parameters in their weight enumerators that were not previously known. We tabulate our results, stating clearly the parameters of the obtained codes and their orders of the automorphism group. We finish with concluding remarks and directions for possible future research. 

\section{Preliminaries}

\subsection{Codes}

We begin by recalling the standard definitions from coding theory. A code $C$
of length $n$ over a Frobenius ring $R$ is a subset of $R^n$. If the code is
a submodule of $R^n$ then we say that the code is linear. Elements of the
code $C$ are called codewords of $C$. Let $\mathbf{x}=(x_1,x_2,\dots,x_n)$
and $\mathbf{y}=(y_1,y_2,\dots,y_n)$ be two elements of $R^n.$ The duality
is understood in terms of the Euclidean inner product, namely:
\begin{equation*}
\langle \mathbf{x},\mathbf{y} \rangle_E=\sum x_iy_i.
\end{equation*}
The dual $C^{\bot}$ of the code $C$ is defined as
\begin{equation*}
C^{\bot}=\{\mathbf{x} \in R^n \ | \ \langle \mathbf{x},\mathbf{y}
\rangle_E=0 \ \text{for all} \ \mathbf{y} \in C\}.
\end{equation*}
We say that $C$ is self-orthogonal if $C \subseteq C^\perp$ and is self-dual
if $C=C^{\bot}.$

An upper bound on the minimum Hamming distance of a binary self-dual code
was given in \cite{Rains1}. Specifically, let $d_{I}(n)$ and $d_{II}(n)$ be the
minimum distance of a Type~I (singly-even) and Type~II (doubly-even) binary code of length $n$,
respectively. Then
\begin{equation*}
d_{II}(n) \leq 4\lfloor \frac{n}{24} \rfloor+4
\end{equation*}
and
\begin{equation*}
d_{I}(n)\leq
\begin{cases}
\begin{matrix}
4\lfloor \frac{n}{24} \rfloor+4 \ \ \ if \ n \not\equiv 22 \pmod{24} \\
4\lfloor \frac{n}{24} \rfloor+6 \ \ \ if \ n \equiv 22 \pmod{24}.%
\end{matrix}%
\end{cases}%
\end{equation*}

Self-dual codes meeting these bounds are called \textsl{extremal}.
Throughout the text, we obtain extremal binary codes of different lengths.
Self-dual codes which are the best possible for a given set of parameters is
said to be optimal. Extremal codes are necessarily optimal but optimal codes
are not necessarily extremal.

\subsection{Special Matrices and Group Rings}

To understand the form of the generator matrices which we define later in this work,
we recall some basic definitions of some special matrices and theory on group rings.

A circulant matrix is one where each row is shifted one element to the right relative to the preceding row. We label the circulant matrix as $A=circ(\alpha_1,\alpha_2\dots , \alpha_n),$ where $\alpha_i$ are ring elements. The transpose of a matrix $A,$ denoted by $A^T,$ is a matrix whose rows are the columns of $A,$ i.e., $A^T_{ij}=A_{ji}.$  A symmetric matrix is a square matrix that is equal to its transpose. A persymmetric matrix is a square matrix which is symmetric with respect to the northeast-to-southwest diagonal. Later in this work, we only consider $2 \times 2$ persymmetric matrices with three independent variables for which we use the following notation:
$$persym(a_1,a_2,a_3)=\begin{pmatrix}
a_1&a_2\\
a_3&a_1
\end{pmatrix}.$$

Let $R$ be a ring, then if $R$ has an identity $1_R,$ we say that $u \in R$ is a unit in $R$ if and only if there exists an element $w \in R$ with $uw=1_R.$While group rings can be given for infinite rings and infinite groups, we
are only concerned with group rings where both the ring and the group are
finite. Let $G$ be a finite group of order $n$, then the group ring $RG$
consists of $\sum_{i=1}^n \alpha_i g_i$, $\alpha_i \in R$, $g_i \in G.$

Addition in the group ring is done by coordinate addition, namely
\begin{equation}
\sum_{i=1}^n \alpha_i g_i +\sum_{i=1}^n \beta_i g_i =\sum_{i=1}^n (\alpha_i
+ \beta_i)g_i.
\end{equation}
The product of two elements in a group ring is given by
\begin{equation}
\left(\sum_{i=1}^n \alpha_i g_i \right)\left(\sum_{j=1}^n \beta_j g_j
\right)= \sum_{i,j} \alpha_i \beta_j g_i g_j.
\end{equation}
It follows that the coefficient of $g_k$ in the product is $\sum_{g_i
g_j=g_k} \alpha_i \beta_j.$

\subsection{The map $\tau_k(v)$ and generator matrices of the form $[I_{kn} \ | \ \tau_k(v)]$}

We now recall the map $\tau_k(v),$ where $v \in M_k(R)G$ and where $M_k(R)$ is a non-commutative Frobenius matrix ring and $G$ is a finite group of order $n,$ that was introduced in \cite{Dougherty3}.

Let $v=A_{g_1}g_1+A_{g_2}g_2+\dots+A_{g_n}g_n \in M_k(R)G,$ that is, each $A_{g_i}$ is a $k \times k$ matrix with entries from the ring $R.$ Define the block matrix $\sigma_k(v) \in (M_{k}(R))_n$ to be

\begin{equation}\label{sigmakv}
\sigma_k(v)=
\begin{pmatrix}
A_{g_1^{-1}g_1} & A_{g_1^{-1}g_2} & A_{g_1^{-1}g_3} & \dots &
A_{g_1^{-1}g_n} \\
A_{g_2^{-1}g_1} & A_{g_2^{-1}g_2} & A_{g_2^{-1}g_3} & \dots &
A_{g_2^{-1}g_n} \\
\vdots & \vdots & \vdots & \vdots & \vdots \\
A_{g_n^{-1}g_1} & A_{g_n^{-1}g_2} & A_{g_n^{-1}g_3} & \dots &
A_{g_n^{-1}g_n}
\end{pmatrix}
.
\end{equation}

We note that the element $v$ is an element of the group matrix ring $M_k(R)G.$

{\bf Construction 1} 
For a given element $v \in M_k(R)G,$ we define the following code over the matrix ring $M_k(R)$:
\begin{equation}
C_k(v)=\langle \sigma_k(v) \rangle.
\end{equation}
Here the code is generated by taking the all left linear combinations of the rows of the matrix with coefficients in $M_k(R).$ 

{\bf Construction 2} 
For a given element $v \in M_k(R)G,$ we define the following  code over the  ring $R$.
Construct the matrix $\tau_k(v)$ by viewing each element in a $k$ by $k$ matrix as an element in the larger matrix.  
\begin{equation}
B_k(v)=\langle \tau_k(v) \rangle.
\end{equation}
Here the code $B_k(v)$ is formed by taking all linear combinations of the rows of the matrix with coefficients in $R$.  
In this case the ring over which the code is defined is commutative so it is both a left linear and right linear code.   

We note that the map $\tau_k(v)$ does not only depend on the choice of the group $G$ and the ring $R,$ but also on the structure of the $k \times k$ matrices. Later in this work, we employ this map and consider generator matrices of the form $[I_{kn} \ | \ \tau_k(v)]$ for groups of order $18$ and for $k=2.$ That is, we consider $2 \times 2$ matrices of different forms. We finish this section with some results on the generator matrices of the form $[I_{kn} \ | \ \tau_k(v)]$ from \cite{Dougherty3}.

\begin{lemma}\label{I|taukv}
Let $G$ be a group of order $n$ and $v=A_1g_1+A_2g_2+\dots +A_ng_n$ be an element of the group matrix ring $M_k(R)G.$ The matrix $[I_{kn}|\tau_k(v)]$ generates a self-dual code over $R$ if and only if $\tau_k(v)\tau_k(v)^T=-I_{kn}$.
\end{lemma}

Recall that the canonical involution $* : RG \rightarrow RG$ on a group ring $RG$ is given by $v^*=\sum_g a_g g^{-1},$ for $v=\sum_g a_g g \in RG.$ Also, recall that there is a connection between $v^*$ and $v$ when we take their images under the map $\sigma,$ given by
\begin{equation}
\sigma(v^*)=\sigma(v)^T.
\end{equation}
The above connection can be extended to the group matrix ring $M_k(R)G.$ Namely, let $* : M_k(R)G\rightarrow M_k(R)G$ be the canonical involution on the group matrix ring $M_k(R)G$ given by $v^*=\sum_g A_gg^{-1},$ for $v=\sum_g A_gg \in M_k(R)G$ where $A_g$ are the $k \times k$ blocks. Then we have the following connection between $v^*$ and $v$ under the map $\tau_k$:
\begin{equation}
\tau_k(v^*)=\tau_k(v)^T.
\end{equation} 

\begin{lemma}\label{taukisringhomomorphism}
Let $R$ be a finite commutative ring. Let $G$ be a group of order $n$ with a fixed listing of its elements. Then the map $\tau_k : v \rightarrow M(R)_{kn}$ is a bijective ring homomorphism.
\end{lemma}
 
Now, combining together Lemma~\ref{I|taukv}, Lemma~\ref{taukisringhomomorphism} and the fact that $\tau_k(v)=-I_{kn}$ if and only if $v=-I_k,$ we get the following corollary.

\begin{corollary}
Let $M_k(R)G$ be a group matrix ring, where $M_k(R)$ is a non-commutative Frobenius matrix ring. For $v \in M_k(R)G,$ the matrix $[I_{kn}|\tau_k(v)]$ generates a self-dual code over $R$ if and only if $vv^*=-I_k.$ In particular $v$ has to be a unit.
\end{corollary}

When we restrict our attention to a matrix ring of characteristic 2, we have that $-I_k=I_k,$ which leads to the following further corollary:

\begin{corollary}
Let $M_k(R)G$ be a group matrix ring, where $M_k(R)$ is a non-commutative Frobenius matrix ring of characteristic 2. Then the matrix $[I_{kn}|\tau_k(v)]$ generates a self-dual code over $R$ if and only if $v$ satisfies $vv^*=I_k,$ namely $v$ is a unitary unit in $M_k(R)G.$ 
\end{corollary}

\section{Generator Matrices}

In this section, we define generator matrices of the form $[I_{2n} \ | \ \tau_2(v)]$ where $v \in M_2(R)G,$ for groups of order $18$ and some $2 \times 2$ matrices. 

\begin{enumerate}
\item[I.] Let $G=\langle x,y \ | \ x^9=y^2=1, x^y=x^{-1} \rangle \cong D_{18}.$ Also, let $v_1=\sum_{i=0}^8 \sum_{j=0}^1 A_{1+i+8j} y^jx^i \in M(R)D_{18}.$
Then:
$$\tau_2(v_1)=\begin{pmatrix}
A&B\\
B&A
\end{pmatrix}$$
where 
$$A=CIRC(A_1,A_2,A_3,\dots,A_9),$$
$$B=REVCIRC(A_{10},A_{11},A_{12},\dots,A_{18}),$$
and where $A_i \in M_2(R).$ Now we define five generator matrices of the following forms:
\begin{itemize}
\item[1.]
\begin{equation}
\mathcal{G}_1=[I_{36} \ | \ \tau_2(v_1)],
\end{equation}
with 
$$A_1=circ(a_1,a_2), A_2=circ(a_3,a_4),$$
$$\dots,$$
$$A_{17}=circ(a_{33},a_{34}), A_{18}=circ(a_{35},a_{36}).$$
\item[2.]
\begin{equation}
\mathcal{G}_1'=[I_{36} \ | \ \tau_2(v_1)],
\end{equation}
with
$$A_1=circ(a_1,a_2), \dots, A_9=circ(a_{17},a_{18}),$$
$$A_{10}=persym(a_{19},a_{20},a_{21}), \dots, A_{18}=persym(a_{43},a_{44},a_{45}).$$
\item[3.]
\begin{equation}
\mathcal{G}_1''=[I_{36} \ | \ \tau_2(v_1)],
\end{equation}
with
$$A_1=circ(a_1,a_2), A_2=persym(a_3,a_4,a_5), A_3=circ(a_6,a_7),$$
$$ \dots,$$
$$A_{7}=circ(a_{16},a_{17}), A_8=persym(a_{18},a_{19},a_{20}), A_9=circ(a_{21},a_{22}),$$
$$A_{10}=persym(a_{23},a_{24},a_{25}), A_{11}=circ(a_{26},a_{27}),$$
$$A_{12}=persym(a_{28},a_{29},a_{30}), A_{13}=circ(a_{31},a_{32}),$$
$$ \dots,$$
$$A_{17}=circ(a_{41},a_{42}), A_{18}=persym(a_{43},a_{44},a_{45}),$$
\item[4.]
\begin{equation}
\mathcal{G}_1'''=[I_{36} \ | \ \tau_2(v_1)],
\end{equation}
with
$$A_1=persym(a_1,a_2,a_3), \dots, A_9=persym(a_{25},a_{26},a_{27}),$$
$$A_{10}=circ(a_{28},a_{29}), \dots, A_{18}=circ(a_{44},a_{45}).$$
\item[5.]
\begin{equation}
\mathcal{G}_1''''=[I_{36} \ | \ \tau_2(v_1)],
\end{equation}
with
$$A_1=persym(a_1,a_2,a_3), \dots, A_2=persym(a_{4},a_{5},a_{6}),$$
$$\dots$$
$$A_{17}=persym(a_{49},a_{50},a_{51}), \dots, A_{18}=persym(a_{52},a_{53},a_{54}).$$
\end{itemize}

\item[II.] Let $G=\langle x,y \ | \ x^9=y^2=1, x^y=x^{-1}\rangle \cong D_{18}.$ Also, let $v_2=\sum_{i=0}^8 \sum_{j=0}^1 A_{1+i+8j}x^iy^j \in M(R)D_{18}.$ Then:
$$\tau_2(v_2)=\begin{pmatrix}
A&B\\
B^T&A^T
\end{pmatrix}$$
where 
$$A=CIRC(A_1,A_2,A_3,\dots,A_9),$$
$$B=CIRC(A_{10},A_{11},A_{12},\dots,A_{18}),$$
and where $A_i \in M_2(R).$ Now we define three generator matrices of the following forms:
\begin{itemize}
\item[1.]
\begin{equation}
\mathcal{G}_2=[I_{36} \ | \ \tau_2(v_2)],
\end{equation}
with  
$$A_1=circ(a_1,a_2), A_2=circ(a_3,a_4),$$
$$\dots,$$
$$A_{17}=circ(a_{33},a_{34}), A_{18}=circ(a_{35},a_{36}).$$
\item[2.]
\begin{equation}
\mathcal{G}_2'=[I_{36} \ | \ \tau_2(v_2)],
\end{equation}
with
$$A_1=persym(a_1,a_2,a_3), \dots, A_9=persym(a_{25},a_{26},a_{27}),$$
$$A_{10}=circ(a_{28},a_{29}), \dots, A_{18}=circ(a_{44},a_{45}).$$
\item[3.]
\begin{equation}
\mathcal{G}_2''=[I_{36} \ | \ \tau_2(v_2)],
\end{equation}
with
$$A_1=persym(a_1,a_2,a_3), \dots, A_2=persym(a_{4},a_{5},a_{6}),$$
$$\dots$$
$$A_{17}=persym(a_{49},a_{50},a_{51}), \dots, A_{18}=persym(a_{52},a_{53},a_{54}).$$
\end{itemize}

\item[III.] Let $G=\langle x \ | \ x^{18}=1 \rangle C_{3,6}.$ Also, let $v_3=\sum_{i=0}^2 \sum_{j=0}^5 A_{1+i+3j}x^{6i+j} \in M(R)C_{18}.$ Then:
$$\tau_2(v_3)=\begin{pmatrix}
A&B&C&D&E&F\\
F'&A&B&C&D&E\\
E'&F'&A&B&C&D\\
D'&E'&F'&A&B&C\\
C'&D'&E'&F'&A&B\\
B'&C'&D'&E'&F'&A
\end{pmatrix}$$
where
$$ A=CIRC(A_{1},A_{2},A_{3}), B=CIRC(A_{4},A_{5},A_{6}),$$
$$B'=CIRC(A_{6},A_{4},A_{5}), C=CIRC(A_{7},A_{8},A_{9}),$$$$C'=CIRC(A_{9},A_{7},A_{8}), D=CIRC(A_{10},A_{11},A_{12}),$$
$$D'=CIRC(A_{12},A_{10}, A_{11}), E=CIRC(A_{13},A_{14},A_{15}),$$ $$E'=CIRC(A_{15},A_{13},A_{14}),  F=CIRC(A_{16},A_{17},A_{18}),$$ $$F'=CIRC(A_{18},A_{16},A_{17}),$$
and where $A_i \in M_2(R).$ Now we define a generator matrix of the following form:
\begin{itemize}
\item[1.]
\begin{equation}
\mathcal{G}_3=[I_{36} \ | \ \tau_2(v_3)],
\end{equation}
with  
$$A_1=circ(a_1,a_2), A_2=circ(a_3,a_4),$$
$$\dots,$$
$$A_{17}=circ(a_{33},a_{34}), A_{18}=circ(a_{35},a_{36}).$$
\end{itemize}

\item[IV.] Let $G=\langle x,y \ | \ x^6=y^3=1, xy=yx \rangle \cong C_3 \times C_6.$ Also, let $v_4=\sum_{i=0}^5 \sum_{j=0}^2 A_{1+i+6j}x^iy^j \in M(R)(C_3 \times C_6).$ Then: 
$$\tau_2(v_4)=\begin{pmatrix}
A&B&C\\
C&A&B\\
B&C&A
\end{pmatrix}$$
where
$$ A=CIRC(A_{1}, A_{2}, \dots, A_{6} ), B=CIRC(A_{7}, A_{8}, \dots, A_{12} ),$$
$$C=CIRC(A_{13}, A_{14}, \dots, A_{18} ),$$
and where $A_i \in M_2(R).$ Now we define two generator matrices of the following forms:
\begin{itemize}
\item[1.]
\begin{equation}
\mathcal{G}_4=[I_{36} \ | \ \tau_2(v_4)],
\end{equation}
with  
$$A_1=circ(a_1,a_2), A_2=circ(a_3,a_4),$$
$$\dots,$$
$$A_{17}=circ(a_{33},a_{34}), A_{18}=circ(a_{35},a_{36}).$$
\item[2.]
\begin{equation}
\mathcal{G}_4'=[I_{36} \ | \ \tau_2(v_4)],
\end{equation}
with
$$A_1=persym(a_1,a_2,a_3), \dots, A_9=persym(a_{25},a_{26},a_{27}),$$
$$A_{10}=circ(a_{28},a_{29}), \dots, A_{18}=circ(a_{44},a_{45}).$$
\end{itemize}

\item[V.] Let $G=\langle x,y \ | \ x^6=y^3=1, xy=yx \rangle \cong C_6 \times C_3.$ Also, let $v_5=\sum_{i=0}^5 \sum_{j=0}^2 A_{1+3i+j}x^iy^j \in M(R)(C_6 \times C_3).$ Then:
$$\tau_2(v_5)=\begin{pmatrix}
A&B&C&D&E&F\\
F&A&B&C&D&E\\
E&F&A&B&C&D\\
D&E&F&A&B&C\\
C&D&E&F&A&B\\
B&C&D&E&F&A
\end{pmatrix}$$
where
$$A=CIRC(A_1,A_2,A_3), B=CIRC(A_4,A_5,A_6),$$
$$\dots ,$$
$$E=CIRC(A_{13},A_{14},A_{15}), F=CIRC(A_{16},A_{17},A_{18}),$$
and where $A_i \in M_2(R).$ Now we define a generator matrix of the following form:
\begin{itemize}
\item[1.]
\begin{equation}
\mathcal{G}_5=[I_{36} \ | \ \tau_2(v_5)],
\end{equation}
with  
$$A_1=circ(a_1,a_2), A_2=circ(a_3,a_4),$$
$$\dots,$$
$$A_{17}=circ(a_{33},a_{34}), A_{18}=circ(a_{35},a_{36}).$$
\end{itemize}
\end{enumerate}

We note that in the above generator matrices, the choices for the $2 \times 2$ matrices represent only a fraction of all the possibilities. There are many more possibilities to consider. One may, for example, consider $2 \times 2$ matrices with four independent variables, however, this would increase the number of calculations.

\section{New Binary Self-Dual Codes of length 72}

In this section, we employ the generator matrices defined in Section~3 and search for binary $[72,36,12]$ self-dual codes.

The possible weight enumerators for a Type~I $[72,36,12]$ codes are as follows (\cite{Dougherty4}):
$$W_{72,1}=1+2\beta y^{12}+(8640-64\gamma)y^{14}+(124281-24\beta+384\gamma)y^{16}+\dots$$
$$W_{72,2}=1+2 \beta y^{12}+(7616-64 \gamma)y^{14}+(134521-24 \beta+384 \gamma)y^{16}+\dots$$
where $\beta$ and $\gamma$ are parameters.
The possible weight enumerators for Type~II $[72,36,12]$ codes are (\cite{Dougherty4}):
$$1+(4398+\alpha)y^{12}+(197073-12\alpha)y^{16}+(18396972+66\alpha)y^{20}+\dots $$
where $\alpha$ is a parameter.

Many codes for different values of $\alpha$, $\beta$ and $\gamma$ have been constructed in \cite{Bouyukliev1, Dontcheva1, Dougherty4, Dougherty5, Dougherty3, Gulliver1, Yankov2, Kaya1, Korban1, Yildiz1, Yankov1, Zhdanov1, Zhdanov2}. For an up-to-date list of all known Type~I and Type~II binary self-dual codes with parameters $[72,36,12]$ please see \cite{selfdual72}.

We now split the remaining of this section and tabulate our findings according to the generator matrix we employ. We only list codes with parameters in their weight enumerators that were not known in the literature before. All the upcoming computational results were obtained by performing searches using a particular algorithm technique (see \cite{Korban1} for details) in the software package MAGMA (\cite{MAGMA}).

\begin{enumerate}
\item[1.] Generator matrices $\mathcal{G}_1, \mathcal{G}_1', \mathcal{G}_1'', \mathcal{G}_1'''$ and $\mathcal{G}_1''''$

In the generator matrix $\mathcal{G}_1,$ the matrix $\tau_2(v_1)$ is fully defined by the first row, for this reason, we only list the first row of the matrices $A$ and $B$ which we label as $r_{A}$ and $r_{B}$ respectively.

\newpage
\begin{table}[h!]
\caption{New Type I $[72,36,12]$ Codes from $\mathcal{G}_1$ and $R=\mathbb{F}_2$}
\resizebox{0.65\textwidth}{!}{\begin{minipage}{\textwidth}
\centering
\begin{tabular}{ccccccc}
\hline
      & Type       & $r_A$                                   & $r_B$                                   & $\gamma$ & $\beta$ & $|Aut(C_i)|$ \\ \hline
$C_1$ & $W_{72,1}$ & $(0,0,0,0,0,1,1,0,0,0,0,1,1,1,0,0,0,0)$ & $(0,1,1,0,0,1,0,1,1,1,1,0,1,1,1,0,1,1)$ & $0$      & $93$   & $36$         \\ \hline
$C_2$ & $W_{72,1}$ & $(0,1,1,0,1,0,0,1,0,0,0,0,1,0,1,0,0,0)$ & $(1,0,0,1,1,0,0,0,1,1,0,0,1,1,0,0,0,0)$ & $0$      & $111$   & $36$         \\ \hline
$C_3$ & $W_{72,1}$ & $(0,1,0,1,0,1,1,0,1,0,0,0,0,1,1,1,1,1)$ & $(1,1,0,0,1,1,0,0,0,1,0,0,1,0,0,0,1,0)$ & $0$      & $132$   & $36$         \\ \hline
$C_4$ & $W_{72,1}$ & $(0,0,0,0,1,1,0,1,1,0,0,1,1,1,0,0,1,0)$ & $(0,1,0,0,1,0,1,0,1,0,0,0,1,0,0,0,0,0)$ & $0$      & $138$   & $36$         \\ \hline
$C_5$ & $W_{72,1}$ & $(0,0,1,1,0,1,0,1,0,1,1,0,0,1,0,1,0,0)$ & $(1,0,0,0,1,1,0,1,0,1,1,1,0,0,0,0,1,1)$ & $0$      & $144$   & $36$         \\ \hline
$C_6$ & $W_{72,1}$ & $(0,1,0,0,1,1,1,0,1,0,0,1,0,0,0,1,0,0)$ & $(0,1,1,1,1,0,0,1,1,1,1,1,0,0,0,0,0,1)$ & $0$      & $150$   & $36$         \\ \hline
$C_{7}$ & $W_{72,1}$ & $(0, 1, 0, 0, 1, 0, 0, 0, 0, 0, 1, 1, 0, 0, 0, 1, 1, 1)$ & $(1, 0, 1, 0, 0, 0, 1, 1, 1, 0, 1, 1, 0, 1, 1, 0, 0, 1)$ & $0$      & $174$   & $36$         \\ \hline
$C_{8}$ & $W_{72,1}$ & $(0,1,1,1,1,1,0,1,0,0,0,1,1,1,0,0,0,0)$ & $(1,0,0,1,0,1,1,1,0,0,0,1,1,0,1,0,0,0)$ & $0$      & $198$   & $36$         \\ \hline
$C_9$ & $W_{72,1}$ & $(0, 0, 1, 1, 0, 0, 0, 1, 1, 0, 1, 0, 0, 0, 0, 1, 1, 0)$ & $(0, 1, 1, 1, 0, 1, 1, 1, 1, 1, 0, 0, 0, 1, 0, 0, 1, 0)$ & $0$      & $309$   & $36$         \\ \hline
$C_{10}$ & $W_{72,1}$ & $(0,1,1,1,0,1,1,1,1,1,0,0,1,1,0,1,0,1)$ & $(0,1,1,0,1,1,0,1,0,0,1,0,1,0,1,1,0,0)$ & $0$      & $345$   & $36$         \\ \hline
$C_{11}$ & $W_{72,1}$ & $(1,1,0,1,1,0,1,0,0,0,1,1,0,0,0,1,0,0)$ & $(1,1,0,1,1,0,0,1,1,1,1,0,1,1,1,1,0,1)$ & $0$      & $366$   & $36$         \\ \hline
$C_{12}$ & $W_{72,1}$ & $(1,0,1,1,0,0,0,0,1,1,1,1,0,1,1,1,0,1)$ & $(0,0,0,1,1,0,0,1,1,0,0,0,0,1,0,0,0,1)$ & $0$      & $378$   & $36$         \\ \hline
$C_{13}$ & $W_{72,1}$ & $(1,0,1,0,1,0,1,1,0,1,0,0,0,1,0,1,0,0)$ & $(1,0,1,0,1,1,0,0,0,0,1,0,1,1,1,1,0,0)$ & $0$      & $411$   & $36$         \\ \hline
$C_{14}$ & $W_{72,1}$ & $(1,1,1,0,1,1,0,1,1,1,1,0,0,1,0,1,1,0)$ & $(0,1,0,1,1,1,1,1,1,1,1,1,1,1,1,0,0,0)$ & $0$      & $444$   & $36$         \\ \hline
$C_{15}$ & $W_{72,1}$ & $(1,0,1,0,0,0,1,0,1,0,1,1,0,1,0,0,1,0)$ & $(0,1,0,1,1,1,1,1,1,1,1,0,0,0,0,0,0,0)$ & $0$      & $453$   & $36$         \\ \hline
$C_{16}$ & $W_{72,1}$ & $(0,0,0,1,1,1,1,1,1,1,0,1,0,1,0,0,1,0)$ & $(0,0,0,1,1,0,0,1,1,0,0,0,0,0,0,1,1,1)$ & $18$      & $228$   & $36$         \\ \hline
$C_{17}$ & $W_{72,1}$ & $(1,0,0,1,0,0,1,1,1,1,0,1,1,0,0,1,0,1)$ & $(1,0,0,0,1,1,0,0,0,0,0,1,0,1,1,1,0,0)$ & $18$      & $243$   & $36$         \\ \hline
$C_{18}$ & $W_{72,1}$ & $(1,1,1,0,0,1,1,1,1,0,0,1,1,0,1,1,1,1)$ & $(1,1,1,0,1,1,0,0,0,0,1,0,0,0,0,1,0,1)$ & $18$      & $252$   & $36$         \\ \hline
$C_{19}$ & $W_{72,1}$ & $(1,0,0,1,1,1,1,0,0,1,0,0,0,0,0,1,1,0)$ & $(1,1,0,0,0,0,1,1,0,0,1,0,1,0,1,0,1,1)$ & $18$      & $255$   & $36$         \\ \hline
$C_{20}$ & $W_{72,1}$ & $(0,1,1,0,0,0,0,1,0,0,1,1,0,0,1,0,1,1)$ & $(1,1,1,0,1,0,0,1,1,1,0,0,0,0,1,0,0,1)$ & $18$      & $267$   & $36$         \\ \hline
$C_{21}$ & $W_{72,1}$ & $(1, 0, 1, 0,  0, 1, 1, 1, 0, 1, 1, 1, 1, 0, 1, 1, 0, 1)$ & $(1, 1, 0, 0, 1, 0, 0, 0, 1, 0, 0, 1, 1, 1, 0, 0, 1, 1 )$ & $18$      & $282$   & $36$         \\ \hline
$C_{22}$ & $W_{72,1}$ & $(0,1,1,1,0,1,0,0,0,1,0,1,0,0,1,1,1,0)$ & $(1,1,1,1,1,1,0,1,1,1,1,0,1,0,0,1,0,0)$ & $18$      & $291$   & $36$         \\ \hline
$C_{23}$ & $W_{72,1}$ & $(1,1,0,1,0,1,0,1,0,1,1,0,0,0,1,0,1,1)$ & $(1,1,0,1,0,0,1,0,0,0,1,0,0,0,0,0,1,1)$ & $18$      & $294$   & $36$         \\ \hline
$C_{24}$ & $W_{72,1}$ & $(1,1,1,0,0,1,0,1,0,1,0,0,0,1,0,0,0,0)$ & $(0,1,1,1,1,0,0,0,0,0,0,1,0,1,1,1,1,1)$ & $18$      & $303$   & $36$         \\ \hline
$C_{25}$ & $W_{72,1}$ & $(0,1,0,1,1,1,1,0,1,0,0,0,1,0,1,1,1,0)$ & $(1,1,1,0,0,0,0,0,1,1,0,1,1,0,0,0,0,0)$ & $18$      & $312$   & $36$         \\ \hline
$C_{26}$ & $W_{72,1}$ & $(0,1,0,0,1,0,1,0,0,1,0,0,1,0,0,1,1,1)$ & $(0,1,1,1,1,1,0,0,1,0,0,0,0,0,1,1,0,1)$ & $18$      & $318$   & $36$         \\ \hline
$C_{27}$ & $W_{72,1}$ & $(0,0,1,1,1,0,1,0,1,1,0,1,1,1,0,1,1,0)$ & $(0,1,0,1,0,1,0,0,1,0,1,1,0,0,1,1,1,1)$ & $18$      & $321$   & $36$         \\ \hline
$C_{28}$ & $W_{72,1}$ & $(1,0,1,0,0,1,1,1,0,1,1,1,0,1,1,1,1,0)$ & $(0,1,0,0,1,0,1,0,1,1,0,0,0,0,0,0,0,0)$ & $18$      & $330$   & $36$         \\ \hline
$C_{29}$ & $W_{72,1}$ & $(0,0,1,0,0,1,1,0,0,1,1,1,0,1,0,0,1,0)$ & $(0,1,0,0,1,0,1,0,1,1,0,0,0,0,0,0,0,0)$ & $18$      & $333$   & $36$         \\ \hline
$C_{30}$ & $W_{72,1}$ & $(0,1,1,1,1,0,1,1,0,0,1,0,1,0,1,1,0,0)$ & $(0,1,1,0,0,1,1,1,1,0,1,1,0,0,1,1,0,1)$ & $18$      & $339$   & $36$         \\ \hline
$C_{31}$ & $W_{72,1}$ & $(0,0,0,1,0,0,1,0,0,0,1,0,1,0,0,1,1,0)$ & $(1,0,0,1,0,0,1,1,1,1,0,0,1,1,0,1,1,1)$ & $18$      & $348$   & $36$         \\ \hline
$C_{32}$ & $W_{72,1}$ & $(1,1,0,0,0,1,1,0,1,0,1,0,1,0,1,0,0,0)$ & $(1,1,1,1,0,1,1,1,1,1,0,0,1,1,0,1,1,0)$ & $18$      & $351$   & $36$         \\ \hline
$C_{33}$ & $W_{72,1}$ & $(1,0,0,1,1,1,1,1,1,1,0,1,1,0,1,0,0,1)$ & $(1,0,0,0,1,1,1,0,1,1,0,0,1,1,0,0,1,0)$ & $18$      & $360$   & $36$         \\ \hline
$C_{34}$ & $W_{72,1}$ & $(0,1,1,0,1,0,0,0,1,0,1,1,1,0,0,0,1,0)$ & $(0,0,0,0,1,0,0,0,0,1,0,1,0,0,0,0,1,1)$ & $18$      & $363$   & $36$         \\ \hline
$C_{35}$ & $W_{72,1}$ & $(1,0,1,1,1,1,1,0,1,1,1,1,0,1,0,1,0,1)$ & $(1,1,1,0,0,0,0,1,0,1,0,0,0,0,1,1,1,0)$ & $18$      & $366$   & $36$         \\ \hline
$C_{36}$ & $W_{72,1}$ & $(0, 1, 0, 1, 0, 1, 1, 1, 0, 1, 1, 1, 0, 1, 0, 0, 0, 1)$ & $(0, 0, 1, 1, 0, 0, 1, 0, 0, 0, 0, 1, 0, 1, 0, 0, 1, 1)$ & $18$      & $369$   & $36$         \\ \hline
$C_{37}$ & $W_{72,1}$ & $(0,1,0,0,1,0,0,0,1,1,1,1,1,1,1,0,0,0)$ & $(1,0,1,0,0,1,1,0,1,0,0,0,1,0,0,0,1,1)$ & $18$      & $372$   & $36$         \\ \hline
$C_{38}$ & $W_{72,1}$ & $(0, 1, 0, 0, 0, 0, 0, 1, 1, 0, 0, 0, 0, 0, 0, 1, 1, 1)$ & $(1, 0, 1, 0, 0, 0, 0, 1, 0, 0, 0, 0, 1, 1, 1, 0, 1, 0)$ & $18$      & $381$   & $36$         \\ \hline
$C_{39}$ & $W_{72,1}$ & $(1, 0, 0, 1, 1, 0, 1, 0, 0, 0, 0, 1, 0, 0, 0, 0, 1, 0)$ & $(1, 1, 0, 1, 0, 0, 1, 0, 1, 1, 1, 1, 0, 0, 1, 1, 1, 0)$ & $18$      & $384$   & $36$         \\ \hline
$C_{40}$ & $W_{72,1}$ & $(0,0,1,0,0,0,0,0,0,1,0,0,0,1,0,1,1,0)$ & $(1,1,0,1,1,1,1,1,0,0,0,1,0,1,0,1,1,1)$ & $18$      & $390$   & $36$         \\ \hline
$C_{41}$ & $W_{72,1}$ & $(0,1,0,0,1,0,0,0,1,1,0,0,1,1,0,1,1,0)$ & $(0,0,0,1,0,0,1,0,1,0,1,1,1,1,1,0,0,1)$ & $18$      & $399$   & $36$         \\ \hline
$C_{42}$ & $W_{72,1}$ & $(1,1,1,0,0,1,0,0,1,1,1,1,1,0,0,0,1,0)$ & $(0,0,0,1,0,1,1,0,1,0,0,0,0,0,0,1,1,1)$ & $18$      & $402$   & $36$         \\ \hline
$C_{43}$ & $W_{72,1}$ & $(0,0,1,0,0,0,1,1,0,1,1,0,1,0,0,1,0,0)$ & $(1,1,1,0,0,1,0,0,1,1,0,0,0,1,1,1,0,1)$ & $18$      & $408$   & $36$         \\ \hline
$C_{44}$ & $W_{72,1}$ & $(0,0,1,1,1,1,0,1,1,0,0,0,0,1,0,1,1,1)$ & $(0,1,1,1,1,0,0,1,0,1,1,1,1,1,0,1,0,0)$ & $18$      & $411$   & $36$         \\ \hline
$C_{45}$ & $W_{72,1}$ & $(0,0,1,1,1,0,0,0,1,0,0,0,1,1,1,0,1,0)$ & $(1,0,1,0,1,1,1,0,1,1,1,1,1,1,1,0,0,1)$ & $18$      & $414$   & $36$         \\ \hline
$C_{46}$ & $W_{72,1}$ & $(0, 1, 1, 0, 1, 0, 1, 0, 1, 1, 1, 1, 0, 1, 0, 0, 0, 0 )$ & $(0, 0, 1, 0, 1, 1, 0, 1, 0, 0, 1, 0, 0, 1, 0, 0, 1, 1)$ & $18$      & $417$   & $36$         \\ \hline
$C_{47}$ & $W_{72,1}$ & $(0,1,1,1,0,0,1,0,1,1,0,1,0,0,0,0,1,0)$ & $(0,0,1,1,0,1,0,0,1,0,0,1,1,0,1,0,1,1)$ & $18$      & $423$   & $36$         \\ \hline
$C_{48}$ & $W_{72,1}$ & $(1, 0, 1, 1, 0, 0, 1, 1, 1, 1, 1, 0, 0, 0, 0, 1, 1, 0)$ & $( 1, 1, 0, 1, 0, 0, 0, 1, 1, 0, 0, 0, 0, 0, 1, 0, 1, 0)$ & $18$      & $426$   & $36$         \\ \hline
$C_{49}$ & $W_{72,1}$ & $(1,1,0,0,1,0,1,0,1,0,1,1,1,1,0,0,1,1)$ & $(0,1,1,1,0,0,0,1,1,0,1,0,1,0,1,1,1,0)$ & $18$      & $438$   & $36$         \\ \hline
$C_{50}$ & $W_{72,1}$ & $(0,0,0,1,0,0,1,1,0,1,1,0,1,0,0,1,0,0)$ & $(1,1,1,0,0,1,0,0,1,1,0,0,0,1,1,1,0,1)$ & $18$      & $444$   & $36$         \\ \hline
\end{tabular}
\end{minipage}}
\end{table}

\begin{table}[h!]
\resizebox{0.65\textwidth}{!}{\begin{minipage}{\textwidth}
\centering
\begin{tabular}{ccccccc}
\hline
      & Type       & $r_A$                                   & $r_B$                                   & $\gamma$ & $\beta$ & $|Aut(C_i)|$ \\ \hline
$C_{51}$ & $W_{72,1}$ & $(1,0,0,0,0,1,1,1,1,0,0,1,1,0,1,1,1,1)$ & $(0,1,0,0,1,0,0,1,1,1,1,1,1,1,0,0,0,1)$ & $18$      & $450$   & $36$         \\ \hline
$C_{52}$ & $W_{72,1}$ & $(1,0,0,1,0,1,0,0,1,1,1,1,1,1,0,0,0,0)$ & $(1,0,1,1,1,1,0,1,1,0,1,0,0,1,1,1,0,1)$ & $18$      & $462$   & $36$         \\ \hline
$C_{53}$ & $W_{72,1}$ & $(0,0,0,0,0,1,0,1,1,1,1,0,1,1,1,0,0,1)$ & $(1,1,0,0,0,0,0,0,1,0,1,0,1,0,1,1,1,0)$ & $18$      & $471$   & $36$         \\ \hline
$C_{54}$ & $W_{72,1}$ & $(0,0,1,0,0,1,1,0,0,1,1,1,1,0,0,1,0,0)$ & $(1,1,0,0,1,1,0,1,1,0,0,1,1,1,1,1,1,1)$ & $18$      & $474$   & $36$         \\ \hline
$C_{55}$ & $W_{72,1}$ & $(0,1,1,0,1,1,0,1,1,1,1,0,1,0,0,1,1,1)$ & $(1,0,0,0,1,1,0,0,0,1,1,0,1,1,1,1,0,0)$ & $18$      & $480$   & $36$         \\ \hline
$C_{56}$ & $W_{72,1}$ & $(0,0,1,0,1,0,0,0,1,1,1,1,1,1,1,0,0,0)$ & $(0,1,0,1,0,1,1,0,0,0,0,0,1,1,1,0,1,0)$ & $18$      & $486$   & $36$         \\ \hline
$C_{57}$ & $W_{72,1}$ & $(1,0,0,0,0,1,0,1,0,1,0,0,0,1,1,1,0,0)$ & $(1,1,1,1,0,1,0,0,0,0,1,0,0,1,0,1,1,1)$ & $18$      & $489$   & $36$         \\ \hline
$C_{58}$ & $W_{72,1}$ & $(0,0,1,0,1,0,0,0,1,1,1,0,0,0,1,1,0,0)$ & $(1,1,0,1,1,0,1,0,1,0,0,1,1,0,1,1,0,0)$ & $18$      & $498$   & $36$         \\ \hline
$C_{59}$ & $W_{72,1}$ & $(1,1,1,0,0,1,0,0,0,0,1,1,0,1,1,0,1,0)$ & $(1,1,0,0,1,0,0,0,1,0,1,0,0,0,1,0,1,1)$ & $18$      & $507$   & $36$         \\ \hline
$C_{60}$ & $W_{72,1}$ & $(1,0,1,1,0,0,0,0,0,0,0,1,1,1,0,1,1,1)$ & $(0,1,0,0,0,1,1,0,1,0,1,0,0,1,0,0,1,1)$ & $18$      & $516$   & $36$         \\ \hline
$C_{61}$ & $W_{72,1}$ & $(1,0,1,0,1,0,1,1,0,1,1,1,0,0,1,0,0,1)$ & $(0,0,0,0,1,0,0,0,1,0,1,1,1,0,1,1,0,0)$ & $18$      & $525$   & $36$         \\ \hline
$C_{62}$ & $W_{72,1}$ & $(1,0,1,0,1,0,0,1,0,1,1,1,0,0,0,0,0,1)$ & $(0,1,1,0,1,1,1,1,1,0,0,0,0,0,1,1,0,0)$ & $18$      & $540$   & $36$         \\ \hline
$C_{63}$ & $W_{72,1}$ & $(0,0,0,1,0,0,0,0,1,1,1,0,1,0,0,1,0,1)$ & $(0,0,0,0,0,1,1,0,0,0,1,1,1,0,0,0,1,0)$ & $36$      & $393$   & $36$         \\ \hline
$C_{64}$ & $W_{72,1}$ & $(1,1,1,1,0,1,0,0,0,0,1,0,1,0,1,1,1,1)$ & $(0,0,0,0,0,1,1,0,1,0,0,1,0,1,0,1,0,0)$ & $36$      & $399$   & $36$         \\ \hline
$C_{65}$ & $W_{72,1}$ & $(1,1,0,1,0,0,0,0,0,1,0,1,1,0,0,0,0,0)$ & $(1,1,0,1,1,1,0,0,0,1,1,1,1,0,0,1,0,1)$ & $36$      & $402$   & $36$         \\ \hline
$C_{66}$ & $W_{72,1}$ & $(1,0,1,0,0,1,0,0,1,1,0,1,0,1,0,1,1,1)$ & $(1,0,0,0,0,1,0,0,0,0,1,1,1,0,0,0,1,1)$ & $36$      & $444$   & $36$         \\ \hline
$C_{67}$ & $W_{72,1}$ & $(0,1,1,0,0,1,1,1,0,0,1,1,1,0,1,1,1,1)$ & $(0,0,1,0,0,0,0,1,1,0,0,1,0,0,1,0,0,0)$ & $36$      & $453$   & $72$         \\ \hline
$C_{68}$ & $W_{72,1}$ & $(1, 1, 0, 0, 1, 0, 0, 1, 0, 0, 0, 0, 0, 0, 0, 1, 1, 1)$ & $(0, 0, 1, 0, 1, 0, 0, 1, 1, 1, 1, 0, 0, 1, 1, 1, 1, 0)$ & $36$      & $462$   & $36$         \\ \hline
$C_{69}$ & $W_{72,1}$ & $(0,1,1,1,0,1,0,0,1,1,1,1,1,0,1,0,0,1)$ & $(1,1,0,1,0,1,0,0,0,0,0,1,1,1,1,0,1,1)$ & $36$      & $477$   & $36$         \\ \hline
$C_{70}$ & $W_{72,1}$ & $(0, 0, 0, 0, 0, 1, 0, 1, 1, 1, 0, 1, 1, 0, 1, 1, 1, 1)$ & $(0, 0, 0, 1, 0, 1, 1, 0, 1, 1, 0, 0, 1, 0, 0, 0, 0, 1 )$ & $36$      & $489$   & $36$         \\ \hline
$C_{71}$ & $W_{72,1}$ & $(1,1,0,1,1,0,1,0,0,1,1,1,0,1,0,1,0,0)$ & $(1,1,0,0,1,0,0,1,0,1,0,0,0,0,0,0,1,1)$ & $36$      & $507$   & $36$         \\ \hline
$C_{72}$ & $W_{72,1}$ & $(0,0,0,0,0,0,1,0,0,1,1,0,1,1,0,1,1,0)$ & $(0,0,0,1,1,0,0,0,0,1,1,1,1,1,1,1,0,1)$ & $36$      & $516$   & $36$         \\ \hline
$C_{73}$ & $W_{72,1}$ & $(1, 1, 0, 0, 1, 0, 1, 1, 1, 1, 0, 0, 0, 0, 1, 0, 1, 0)$ & $(1, 0, 1, 0, 0, 1, 1, 1, 0, 0, 0, 0, 1, 0, 0, 1, 0, 1)$ & $36$      & $525$   & $36$         \\ \hline
$C_{74}$ & $W_{72,1}$ & $(1, 1, 1, 1, 0, 0, 1, 0, 0, 1, 1, 1, 1, 1, 1, 1, 1, 0)$ & $(1, 0, 0, 0, 1, 0, 0, 0, 0, 1, 0, 1, 1, 0, 1, 1, 0, 1)$ & $36$      & $534$   & $36$         \\ \hline
$C_{75}$ & $W_{72,1}$ & $(0,1,1,0,1,1,0,0,0,1,1,1,0,1,1,1,0,0)$ & $(0,0,1,1,0,0,1,0,1,0,0,1,0,1,0,0,1,0)$ & $36$      & $582$   & $36$         \\ \hline
$C_{76}$ & $W_{72,1}$ & $(0,0,0,1,1,1,1,0,0,1,0,0,0,0,0,1,1,0)$ & $(1,1,1,0,0,0,1,1,0,0,1,1,1,0,1,0,0,1)$ & $36$      & $588$   & $36$         \\ \hline
$C_{77}$ & $W_{72,1}$ & $(1,1,1,0,0,0,0,0,0,0,0,1,1,0,0,1,1,0)$ & $(0,1,0,0,0,0,1,0,1,1,1,0,1,1,1,1,1,0)$ & $36$      & $600$   & $36$         \\ \hline
$C_{78}$ & $W_{72,1}$ & $(0,0,0,1,1,1,0,0,0,0,0,1,1,0,0,1,1,0)$ & $(1,1,0,0,1,0,1,1,0,1,0,0,1,1,1,0,1,0)$ & $36$      & $606$   & $36$         \\ \hline
$C_{79}$ & $W_{72,1}$ & $(0, 0, 1, 1, 1, 1, 0, 1, 1, 0, 1, 0, 1, 1, 0, 1, 1, 1)$ & $(1, 1, 0, 1, 0, 1, 0, 0, 1, 1, 0, 1, 0, 1, 0, 0, 0, 1)$ & $36$      & $624$   & $36$         \\ \hline
$C_{80}$ & $W_{72,1}$ & $(0,0,1,0,0,0,0,0,1,1,1,0,1,0,0,1,0,0)$ & $(1,1,1,0,0,1,0,1,1,1,1,0,0,0,1,1,0,1)$ & $36$      & $663$   & $36$         \\ \hline
$C_{81}$ & $W_{72,1}$ & $(0,1,1,1,0,0,0,1,1,0,0,1,1,0,1,1,0,1)$ & $(0,1,1,0,0,0,0,0,1,1,0,0,1,1,0,0,1,0)$ & $54$      & $651$   & $36$         \\ \hline
$C_{82}$ & $W_{72,1}$ & $(0,0,1,0,1,1,1,1,1,0,0,0,0,0,1,1,0,1)$ & $(0,1,0,0,1,0,0,1,0,1,0,0,1,1,1,0,1,0)$ & $54$      & $657$   & $36$         \\ \hline
\end{tabular}
\end{minipage}}
\end{table}
\newpage
In the generator matrix $\mathcal{G}_1',$ the matrix $\tau_2(v_1)$ is fully defined by the $2 \times 2$ matrices in the first row- some of them are circulant and some of them are persymmetric. For this reason, we only list the first row of the matrices $A_1,A_2,A_3,\dots,A_{18}$ which we label as $r_{A_1},r_{A_2},r_{A_3},\dots,r_{A_{18}}$ respectively. If the matrix $A_i$ is circulant, we only list the first row of such matrix and if the matrix $A_i$ is persymmetric, we only list the three variables that correspond to such matrix.

\begin{table}[h!]
\caption{New Type~I $[72,36,12]$ Codes from $\mathcal{G}_1'$ and $R=\mathbb{F}_2$}
\resizebox{0.65\textwidth}{!}{\begin{minipage}{\textwidth}
\centering
\begin{tabular}{cccccccccccccc}
\hline
                          & Type                        & $r_{A_1}$    & $r_{A_2}$    & $r_{A_3}$    & $r_{A_4}$    & $r_{A_5}$    & $r_{A_6}$    & $r_{A_7}$    & $r_{A_8}$    & $r_{A_9}$    & $\gamma$              & $\beta$                & $|Aut(C_i)|$          \\ \hline
\multirow{3}{*}{$C_{83}$} & \multirow{3}{*}{$W_{72,1}$} & $(1,0)$    & $(0,1)$    & $(1,0)$    & $(1,0)$    & $(0,0)$    & $(1,1)$    & $(1,0)$    & $(0,0)$    & $(1,1)$    & \multirow{3}{*}{$0$} & \multirow{3}{*}{$282$} & \multirow{3}{*}{$18$} \\
                          &                             & $r_{A_{10}}$ & $r_{A_{11}}$ & $r_{A_{12}}$ & $r_{A_{13}}$ & $r_{A_{14}}$ & $r_{A_{15}}$ & $r_{A_{16}}$ & $r_{A_{17}}$ & $r_{A_{18}}$ &                       &                        &                       \\
                          &                             & $(1,0,1)$      & $(0,1,1)$      & $(0,0,0)$      & $(0,0,0)$      & $(1,0,1)$      & $(0,0,1)$      & $(1,1,0)$      & $(1,1,0)$      & $(0,1,0)$      &                       &                        &                       \\ \hline
                        
\end{tabular}
\end{minipage}}
\end{table}

\begin{table}[h!]
\resizebox{0.65\textwidth}{!}{\begin{minipage}{\textwidth}
\centering
\begin{tabular}{cccccccccccccc}
\hline
                          & Type                        & $r_{A_1}$    & $r_{A_2}$    & $r_{A_3}$    & $r_{A_4}$    & $r_{A_5}$    & $r_{A_6}$    & $r_{A_7}$    & $r_{A_8}$    & $r_{A_9}$    & $\gamma$              & $\beta$                & $|Aut(C_i)|$          \\ \hline
\multirow{3}{*}{$C_{84}$} & \multirow{3}{*}{$W_{72,1}$} & $(0,0)$    & $(0,0)$    & $(1,1)$    & $(0,1)$    & $(0,0)$    & $(1,1)$    & $(0,1)$    & $(0,0)$    & $(0,1)$    & \multirow{3}{*}{$9$} & \multirow{3}{*}{$192$} & \multirow{3}{*}{$18$} \\
                          &                             & $r_{A_{10}}$ & $r_{A_{11}}$ & $r_{A_{12}}$ & $r_{A_{13}}$ & $r_{A_{14}}$ & $r_{A_{15}}$ & $r_{A_{16}}$ & $r_{A_{17}}$ & $r_{A_{18}}$ &                       &                        &                       \\
                          &                             & $(1,0,1)$      & $(0,0,1)$      & $(1,0,0)$      & $(1,1,0)$      & $(1,1,0)$      & $(1,1,1)$      & $(1,0,1)$      & $(0,1,0)$      & $(0,0,0)$      &                       &                        &                       \\ \hline
                        
\end{tabular}
\end{minipage}}
\end{table}

\newpage

\begin{table}[h!]
\resizebox{0.65\textwidth}{!}{\begin{minipage}{\textwidth}
\centering
\begin{tabular}{cccccccccccccc}
\hline
                          & Type                        & $r_{A_1}$    & $r_{A_2}$    & $r_{A_3}$    & $r_{A_4}$    & $r_{A_5}$    & $r_{A_6}$    & $r_{A_7}$    & $r_{A_8}$    & $r_{A_9}$    & $\gamma$              & $\beta$                & $|Aut(C_i)|$          \\ \hline
\multirow{3}{*}{$C_{85}$} & \multirow{3}{*}{$W_{72,1}$} & $(0,1)$    & $(1,0)$    & $(0,0)$    & $(0,1)$    & $(1,0)$    & $(1,0)$    & $(0,0)$    & $(0,1)$    & $(1,1)$    & \multirow{3}{*}{$9$} & \multirow{3}{*}{$210$} & \multirow{3}{*}{$18$} \\
                          &                             & $r_{A_{10}}$ & $r_{A_{11}}$ & $r_{A_{12}}$ & $r_{A_{13}}$ & $r_{A_{14}}$ & $r_{A_{15}}$ & $r_{A_{16}}$ & $r_{A_{17}}$ & $r_{A_{18}}$ &                       &                        &                       \\
                          &                             & $(1,0,1)$      & $(1,0,0)$      & $(1,0,1)$      & $(0,0,1)$      & $(1,1,1)$      & $(1,1,0)$      & $(0,0,1)$      & $(0,1,1)$      & $(1,0,1)$      &                       &                        &                       \\ \hline
                        
\end{tabular}
\end{minipage}}
\end{table}

\begin{table}[h!]
\resizebox{0.65\textwidth}{!}{\begin{minipage}{\textwidth}
\centering
\begin{tabular}{cccccccccccccc}
\hline
                          & Type                        & $r_{A_1}$    & $r_{A_2}$    & $r_{A_3}$    & $r_{A_4}$    & $r_{A_5}$    & $r_{A_6}$    & $r_{A_7}$    & $r_{A_8}$    & $r_{A_9}$    & $\gamma$              & $\beta$                & $|Aut(C_i)|$          \\ \hline
\multirow{3}{*}{$C_{86}$} & \multirow{3}{*}{$W_{72,1}$} & $(0,0)$    & $(1,0)$    & $(1,0)$    & $(1,1)$    & $(1,0)$    & $(1,1)$    & $(0,0)$    & $(1,1)$    & $(1,1)$    & \multirow{3}{*}{$9$} & \multirow{3}{*}{$225$} & \multirow{3}{*}{$18$} \\
                          &                             & $r_{A_{10}}$ & $r_{A_{11}}$ & $r_{A_{12}}$ & $r_{A_{13}}$ & $r_{A_{14}}$ & $r_{A_{15}}$ & $r_{A_{16}}$ & $r_{A_{17}}$ & $r_{A_{18}}$ &                       &                        &                       \\
                          &                             & $(1,1,1)$      & $(0,1,0)$      & $(0,1,0)$      & $(1,1,1)$      & $(1,1,0)$      & $(1,0,1)$      & $(0,1,1)$      & $(0,0,1)$      & $(0,0,1)$      &                       &                        &                       \\ \hline
                        
\end{tabular}
\end{minipage}}
\end{table}

\begin{table}[h!]
\resizebox{0.65\textwidth}{!}{\begin{minipage}{\textwidth}
\centering
\begin{tabular}{cccccccccccccc}
\hline
                          & Type                        & $r_{A_1}$    & $r_{A_2}$    & $r_{A_3}$    & $r_{A_4}$    & $r_{A_5}$    & $r_{A_6}$    & $r_{A_7}$    & $r_{A_8}$    & $r_{A_9}$    & $\gamma$              & $\beta$                & $|Aut(C_i)|$          \\ \hline
\multirow{3}{*}{$C_{87}$} & \multirow{3}{*}{$W_{72,1}$} & $(0,0)$    & $(1,1)$    & $(1,1)$    & $(0,1)$    & $(1,0)$    & $(1,1)$    & $(0,0)$    & $(0,1)$    & $(0,1)$    & \multirow{3}{*}{$9$} & \multirow{3}{*}{$228$} & \multirow{3}{*}{$18$} \\
                          &                             & $r_{A_{10}}$ & $r_{A_{11}}$ & $r_{A_{12}}$ & $r_{A_{13}}$ & $r_{A_{14}}$ & $r_{A_{15}}$ & $r_{A_{16}}$ & $r_{A_{17}}$ & $r_{A_{18}}$ &                       &                        &                       \\
                          &                             & $(0,1,0)$      & $(0,1,1)$      & $(0,0,1)$      & $(1,1,0)$      & $(0,0,0)$      & $(1,1,0)$      & $(0,0,1)$      & $(0,1,1)$      & $(0,0,1)$      &                       &                        &                       \\ \hline
                        
\end{tabular}
\end{minipage}}
\end{table}

\begin{table}[h!]
\resizebox{0.65\textwidth}{!}{\begin{minipage}{\textwidth}
\centering
\begin{tabular}{cccccccccccccc}
\hline
                          & Type                        & $r_{A_1}$    & $r_{A_2}$    & $r_{A_3}$    & $r_{A_4}$    & $r_{A_5}$    & $r_{A_6}$    & $r_{A_7}$    & $r_{A_8}$    & $r_{A_9}$    & $\gamma$              & $\beta$                & $|Aut(C_i)|$          \\ \hline
\multirow{3}{*}{$C_{88}$} & \multirow{3}{*}{$W_{72,1}$} & $(0,1)$    & $(0,0)$    & $(1,1)$    & $(0,1)$    & $(0,0)$    & $(1,1)$    & $(0,0)$    & $(0,1)$    & $(0,1)$    & \multirow{3}{*}{$9$} & \multirow{3}{*}{$255$} & \multirow{3}{*}{$18$} \\
                          &                             & $r_{A_{10}}$ & $r_{A_{11}}$ & $r_{A_{12}}$ & $r_{A_{13}}$ & $r_{A_{14}}$ & $r_{A_{15}}$ & $r_{A_{16}}$ & $r_{A_{17}}$ & $r_{A_{18}}$ &                       &                        &                       \\
                          &                             & $(1,0,0)$      & $(1,1,0)$      & $(0,0,1)$      & $(0,0,0)$      & $(1,1,0)$      & $(1,0,1)$      & $(0,1,1)$      & $(0,0,1)$      & $(1,1,0)$      &                       &                        &                       \\ \hline
                        
\end{tabular}
\end{minipage}}
\end{table}

\begin{table}[h!]
\resizebox{0.65\textwidth}{!}{\begin{minipage}{\textwidth}
\centering
\begin{tabular}{cccccccccccccc}
\hline
                          & Type                        & $r_{A_1}$    & $r_{A_2}$    & $r_{A_3}$    & $r_{A_4}$    & $r_{A_5}$    & $r_{A_6}$    & $r_{A_7}$    & $r_{A_8}$    & $r_{A_9}$    & $\gamma$              & $\beta$                & $|Aut(C_i)|$          \\ \hline
\multirow{3}{*}{$C_{89}$} & \multirow{3}{*}{$W_{72,1}$} & $(1,0)$    & $(1,0)$    & $(0,1)$    & $(0,0)$    & $(0,0)$    & $(0,0)$    & $(1,0)$    & $(1,0)$    & $(0,1)$    & \multirow{3}{*}{$9$} & \multirow{3}{*}{$258$} & \multirow{3}{*}{$18$} \\
                          &                             & $r_{A_{10}}$ & $r_{A_{11}}$ & $r_{A_{12}}$ & $r_{A_{13}}$ & $r_{A_{14}}$ & $r_{A_{15}}$ & $r_{A_{16}}$ & $r_{A_{17}}$ & $r_{A_{18}}$ &                       &                        &                       \\
                          &                             & $(0,0,1)$      & $(0,1,1)$      & $(1,0,1)$      & $(0,0,1)$      & $(1,1,1)$      & $(1,1,0)$      & $(1,1,0)$      & $(0,1,1)$      & $(1,1,0)$      &                       &                        &                       \\ \hline
                        
\end{tabular}
\end{minipage}}
\end{table}

\begin{table}[h!]
\resizebox{0.65\textwidth}{!}{\begin{minipage}{\textwidth}
\centering
\begin{tabular}{cccccccccccccc}
\hline
                          & Type                        & $r_{A_1}$    & $r_{A_2}$    & $r_{A_3}$    & $r_{A_4}$    & $r_{A_5}$    & $r_{A_6}$    & $r_{A_7}$    & $r_{A_8}$    & $r_{A_9}$    & $\gamma$              & $\beta$                & $|Aut(C_i)|$          \\ \hline
\multirow{3}{*}{$C_{90}$} & \multirow{3}{*}{$W_{72,1}$} & $(1,0)$    & $(0,1)$    & $(0,0)$    & $(1,0)$    & $(1,0)$    & $(1,0)$    & $(1,1)$    & $(0,1)$    & $(0,1)$    & \multirow{3}{*}{$9$} & \multirow{3}{*}{$261$} & \multirow{3}{*}{$18$} \\
                          &                             & $r_{A_{10}}$ & $r_{A_{11}}$ & $r_{A_{12}}$ & $r_{A_{13}}$ & $r_{A_{14}}$ & $r_{A_{15}}$ & $r_{A_{16}}$ & $r_{A_{17}}$ & $r_{A_{18}}$ &                       &                        &                       \\
                          &                             & $(0,1,0)$      & $(1,0,1)$      & $(0,1,1)$      & $(0,0,1)$      & $(1,1,0)$      & $(1,1,1)$      & $(1,1,0)$      & $(1,0,1)$      & $(1,1,1)$      &                       &                        &                       \\ \hline
                        
\end{tabular}
\end{minipage}}
\end{table}

\begin{table}[h!]
\resizebox{0.65\textwidth}{!}{\begin{minipage}{\textwidth}
\centering
\begin{tabular}{cccccccccccccc}
\hline
                          & Type                        & $r_{A_1}$    & $r_{A_2}$    & $r_{A_3}$    & $r_{A_4}$    & $r_{A_5}$    & $r_{A_6}$    & $r_{A_7}$    & $r_{A_8}$    & $r_{A_9}$    & $\gamma$              & $\beta$                & $|Aut(C_i)|$          \\ \hline
\multirow{3}{*}{$C_{91}$} & \multirow{3}{*}{$W_{72,1}$} & $(1,0)$    & $(1,0)$    & $(0,1)$    & $(0,0)$    & $(1,1)$    & $(0,1)$    & $(0,1)$    & $(1,1)$    & $(1,1)$    & \multirow{3}{*}{$9$} & \multirow{3}{*}{$270$} & \multirow{3}{*}{$18$} \\
                          &                             & $r_{A_{10}}$ & $r_{A_{11}}$ & $r_{A_{12}}$ & $r_{A_{13}}$ & $r_{A_{14}}$ & $r_{A_{15}}$ & $r_{A_{16}}$ & $r_{A_{17}}$ & $r_{A_{18}}$ &                       &                        &                       \\
                          &                             & $(0,1,0)$      & $(1,0,0)$      & $(1,0,1)$      & $(1,0,1)$      & $(0,0,0)$      & $(1,1,0)$      & $(0,1,0)$      & $(1,1,1)$      & $(1,0,1)$      &                       &                        &                       \\ \hline
                        
\end{tabular}
\end{minipage}}
\end{table}

\begin{table}[h!]
\resizebox{0.65\textwidth}{!}{\begin{minipage}{\textwidth}
\centering
\begin{tabular}{cccccccccccccc}
\hline
                          & Type                        & $r_{A_1}$    & $r_{A_2}$    & $r_{A_3}$    & $r_{A_4}$    & $r_{A_5}$    & $r_{A_6}$    & $r_{A_7}$    & $r_{A_8}$    & $r_{A_9}$    & $\gamma$              & $\beta$                & $|Aut(C_i)|$          \\ \hline
\multirow{3}{*}{$C_{92}$} & \multirow{3}{*}{$W_{72,1}$} & $(0,0)$    & $(0,0)$    & $(1,1)$    & $(1,1)$    & $(1,1)$    & $(1,0)$    & $(1,0)$    & $(1,0)$    & $(0,1)$    & \multirow{3}{*}{$9$} & \multirow{3}{*}{$282$} & \multirow{3}{*}{$18$} \\
                          &                             & $r_{A_{10}}$ & $r_{A_{11}}$ & $r_{A_{12}}$ & $r_{A_{13}}$ & $r_{A_{14}}$ & $r_{A_{15}}$ & $r_{A_{16}}$ & $r_{A_{17}}$ & $r_{A_{18}}$ &                       &                        &                       \\
                          &                             & $(1,1,1)$      & $(0,1,0)$      & $(1,0,1)$      & $(1,0,0)$      & $(0,1,0)$      & $(1,0,1)$      & $(0,1,1)$      & $(1,0,1)$      & $(1,1,0)$      &                       &                        &                       \\ \hline
                        
\end{tabular}
\end{minipage}}
\end{table}

\newpage
\begin{table}[h!]
\resizebox{0.65\textwidth}{!}{\begin{minipage}{\textwidth}
\centering
\begin{tabular}{cccccccccccccc}
\hline
                          & Type                        & $r_{A_1}$    & $r_{A_2}$    & $r_{A_3}$    & $r_{A_4}$    & $r_{A_5}$    & $r_{A_6}$    & $r_{A_7}$    & $r_{A_8}$    & $r_{A_9}$    & $\gamma$              & $\beta$                & $|Aut(C_i)|$          \\ \hline
\multirow{3}{*}{$C_{93}$} & \multirow{3}{*}{$W_{72,1}$} & $(0,1)$    & $(0,0)$    & $(1,1)$    & $(0,0)$    & $(0,1)$    & $(1,1)$    & $(0,0)$    & $(0,1)$    & $(0,1)$    & \multirow{3}{*}{$9$} & \multirow{3}{*}{$318$} & \multirow{3}{*}{$18$} \\
                          &                             & $r_{A_{10}}$ & $r_{A_{11}}$ & $r_{A_{12}}$ & $r_{A_{13}}$ & $r_{A_{14}}$ & $r_{A_{15}}$ & $r_{A_{16}}$ & $r_{A_{17}}$ & $r_{A_{18}}$ &                       &                        &                       \\
                          &                             & $(0,0,1)$      & $(0,1,1)$      & $(0,0,1)$      & $(1,1,0)$      & $(1,0,0)$      & $(1,0,1)$      & $(1,1,0)$      & $(0,0,0)$      & $(1,1,0)$      &                       &                        &                       \\ \hline
                        
\end{tabular}
\end{minipage}}
\end{table}

\begin{table}[h!]
\resizebox{0.65\textwidth}{!}{\begin{minipage}{\textwidth}
\centering
\begin{tabular}{cccccccccccccc}
\hline
                          & Type                        & $r_{A_1}$    & $r_{A_2}$    & $r_{A_3}$    & $r_{A_4}$    & $r_{A_5}$    & $r_{A_6}$    & $r_{A_7}$    & $r_{A_8}$    & $r_{A_9}$    & $\gamma$              & $\beta$                & $|Aut(C_i)|$          \\ \hline
\multirow{3}{*}{$C_{94}$} & \multirow{3}{*}{$W_{72,1}$} & $(1,0)$    & $(0,1)$    & $(1,0)$    & $(0,0)$    & $(1,0)$    & $(0,0)$    & $(1,0)$    & $(1,0)$    & $(1,1)$    & \multirow{3}{*}{$9$} & \multirow{3}{*}{$336$} & \multirow{3}{*}{$18$} \\
                          &                             & $r_{A_{10}}$ & $r_{A_{11}}$ & $r_{A_{12}}$ & $r_{A_{13}}$ & $r_{A_{14}}$ & $r_{A_{15}}$ & $r_{A_{16}}$ & $r_{A_{17}}$ & $r_{A_{18}}$ &                       &                        &                       \\
                          &                             & $(1,0,1)$      & $(1,1,1)$      & $(1,0,1)$      & $(0,0,1)$      & $(0,1,1)$      & $(1,1,0)$      & $(0,0,1)$      & $(1,0,0)$      & $(1,0,1)$      &                       &                        &                       \\ \hline
                        
\end{tabular}
\end{minipage}}
\end{table}

\begin{table}[h!]
\resizebox{0.65\textwidth}{!}{\begin{minipage}{\textwidth}
\centering
\begin{tabular}{cccccccccccccc}
\hline
                          & Type                        & $r_{A_1}$    & $r_{A_2}$    & $r_{A_3}$    & $r_{A_4}$    & $r_{A_5}$    & $r_{A_6}$    & $r_{A_7}$    & $r_{A_8}$    & $r_{A_9}$    & $\gamma$              & $\beta$                & $|Aut(C_i)|$          \\ \hline
\multirow{3}{*}{$C_{95}$} & \multirow{3}{*}{$W_{72,1}$} & $(1,1)$    & $(1,0)$    & $(1,0)$    & $(1,0)$    & $(1,1)$    & $(0,0)$    & $(0,1)$    & $(0,1)$    & $(1,0)$    & \multirow{3}{*}{$18$} & \multirow{3}{*}{$393$} & \multirow{3}{*}{$18$} \\
                          &                             & $r_{A_{10}}$ & $r_{A_{11}}$ & $r_{A_{12}}$ & $r_{A_{13}}$ & $r_{A_{14}}$ & $r_{A_{15}}$ & $r_{A_{16}}$ & $r_{A_{17}}$ & $r_{A_{18}}$ &                       &                        &                       \\
                          &                             & $(0,1,1)$      & $(0,1,0)$      & $(1,0,1)$      & $(1,0,0)$      & $(1,0,1)$      & $(1,0,1)$      & $(0,0,0)$      & $(0,0,1)$      & $(1,0,1)$      &                       &                        &                       \\ \hline
                        
\end{tabular}
\end{minipage}}
\end{table}

In the generator matrix $\mathcal{G}_1'',$ the matrix $\tau_2(v_1)$ is fully defined by the $2 \times 2$ matrices in the first row- some of them are circulant and some of them are persymmetric. For this reason, we only list the first row of the matrices $A_1,A_2,A_3,\dots,A_{18}$ which we label as $r_{A_1},r_{A_2},r_{A_3},\dots,r_{A_{18}}$ respectively. If the matrix $A_i$ is circulant, we only list the first row of such matrix and if the matrix $A_i$ is persymmetric, we only list the three variables that correspond to such matrix.

\begin{table}[h!]
\caption{New Type~I $[72,36,12]$ Codes from $\mathcal{G}_1''$ and $R=\mathbb{F}_2$}
\resizebox{0.65\textwidth}{!}{\begin{minipage}{\textwidth}
\centering
\begin{tabular}{cccccccccccccc}
\hline
                          & Type                        & $r_{A_1}$    & $r_{A_2}$    & $r_{A_3}$    & $r_{A_4}$    & $r_{A_5}$    & $r_{A_6}$    & $r_{A_7}$    & $r_{A_8}$    & $r_{A_9}$    & $\gamma$              & $\beta$                & $|Aut(C_i)|$          \\ \hline
\multirow{3}{*}{$C_{96}$} & \multirow{3}{*}{$W_{72,1}$} & $(0,0,0)$    & $(1,0,1)$    & $(1,1)$    & $(0,0,0)$    & $(1,0)$    & $(0,0,0)$    & $(1,0)$    & $(1,1,0)$    & $(0,0)$    & \multirow{3}{*}{$9$} & \multirow{3}{*}{$252$} & \multirow{3}{*}{$18$} \\
                          &                             & $r_{A_{10}}$ & $r_{A_{11}}$ & $r_{A_{12}}$ & $r_{A_{13}}$ & $r_{A_{14}}$ & $r_{A_{15}}$ & $r_{A_{16}}$ & $r_{A_{17}}$ & $r_{A_{18}}$ &                       &                        &                       \\
                          &                             & $(1,1,0)$      & $(0,1)$      & $(0,1,1)$      & $(1,0)$      & $(1,1,1)$      & $(1,0)$      & $(1,0,1)$      & $(1,0)$      & $(0,0,0)$      &                       &                        &                       \\ \hline
                        
\end{tabular}
\end{minipage}}
\end{table}

In the generator matrix $\mathcal{G}_1''',$ the matrix $\tau_2(v_1)$ is fully defined by the $2 \times 2$ matrices in the first row- some of them are circulant and some of them are persymmetric. For this reason, we only list the first row of the matrices $A_1,A_2,A_3,\dots,A_{18}$ which we label as $r_{A_1},r_{A_2},r_{A_3},\dots,r_{A_{18}}$ respectively. If the matrix $A_i$ is circulant, we only list the first row of such matrix and if the matrix $A_i$ is persymmetric, we only list the three variables that correspond to such matrix.

\begin{table}[h!]
\caption{New Type~I $[72,36,12]$ Codes from $\mathcal{G}_1'''$ and $R=\mathbb{F}_2$}
\resizebox{0.65\textwidth}{!}{\begin{minipage}{\textwidth}
\centering
\begin{tabular}{cccccccccccccc}
\hline
                          & Type                        & $r_{A_1}$    & $r_{A_2}$    & $r_{A_3}$    & $r_{A_4}$    & $r_{A_5}$    & $r_{A_6}$    & $r_{A_7}$    & $r_{A_8}$    & $r_{A_9}$    & $\gamma$              & $\beta$                & $|Aut(C_i)|$          \\ \hline
\multirow{3}{*}{$C_{97}$} & \multirow{3}{*}{$W_{72,1}$} & $(0,1,1)$    & $(1,0,1)$    & $(1,1,0)$    & $(1,1,1)$    & $(1,0,1)$    & $(0,1,0)$    & $(1,0,0)$    & $(1,0,1)$    & $(0,1,0)$    & \multirow{3}{*}{$9$} & \multirow{3}{*}{$291$} & \multirow{3}{*}{$18$} \\
                          &                             & $r_{A_{10}}$ & $r_{A_{11}}$ & $r_{A_{12}}$ & $r_{A_{13}}$ & $r_{A_{14}}$ & $r_{A_{15}}$ & $r_{A_{16}}$ & $r_{A_{17}}$ & $r_{A_{18}}$ &                       &                        &                       \\
                          &                             & $(0,1)$      & $(0,0)$      & $(1,0)$      & $(0,1)$      & $(0,0)$      & $(1,1)$      & $(0,0)$      & $(0,1)$      & $(0,0)$      &                       &                        &                       \\ \hline
                        
\end{tabular}
\end{minipage}}
\end{table}

\newpage

\begin{table}[h!]
\resizebox{0.65\textwidth}{!}{\begin{minipage}{\textwidth}
\centering
\begin{tabular}{cccccccccccccc}
\hline
                          & Type                        & $r_{A_1}$    & $r_{A_2}$    & $r_{A_3}$    & $r_{A_4}$    & $r_{A_5}$    & $r_{A_6}$    & $r_{A_7}$    & $r_{A_8}$    & $r_{A_9}$    & $\gamma$              & $\beta$                & $|Aut(C_i)|$          \\ \hline
\multirow{3}{*}{$C_{98}$} & \multirow{3}{*}{$W_{72,1}$} & $(1,1,0)$    & $(1,0,0)$    & $(0,1,0)$    & $(0,1,0)$    & $(0,0,0)$    & $(1,1,0)$    & $(0,1,0)$    & $(1,1,1)$    & $(0,0,1)$    & \multirow{3}{*}{$9$} & \multirow{3}{*}{$300$} & \multirow{3}{*}{$18$} \\
                          &                             & $r_{A_{10}}$ & $r_{A_{11}}$ & $r_{A_{12}}$ & $r_{A_{13}}$ & $r_{A_{14}}$ & $r_{A_{15}}$ & $r_{A_{16}}$ & $r_{A_{17}}$ & $r_{A_{18}}$ &                       &                        &                       \\
                          &                             & $(0,1)$      & $(0,1)$      & $(0,0)$      & $(0,0)$      & $(1,1)$      & $(1,1)$      & $(0,1)$      & $(1,1)$      & $(1,1)$      &                       &                        &                       \\ \hline
                        
\end{tabular}
\end{minipage}}
\end{table}

In the generator matrix $\mathcal{G}_1'''',$ the matrix $\tau_2(v_1)$ is fully defined by the $2 \times 2$ matrices in the first row that are all persymmetric. For this reason, we only list the first row of the matrices $A_1,A_2,A_3,\dots,A_{18}$ which we label as $r_{A_1},r_{A_2},r_{A_3},\dots,r_{A_{18}}$ respectively. We only list the three variables that correspond to such matrix.

\begin{table}[h!]
\caption{New Type~I $[72,36,12]$ Codes from $\mathcal{G}_1''''$ and $R=\mathbb{F}_2$}
\resizebox{0.65\textwidth}{!}{\begin{minipage}{\textwidth}
\centering
\begin{tabular}{cccccccccccccc}
\hline
                          & Type                        & $r_{A_1}$    & $r_{A_2}$    & $r_{A_3}$    & $r_{A_4}$    & $r_{A_5}$    & $r_{A_6}$    & $r_{A_7}$    & $r_{A_8}$    & $r_{A_9}$    & $\gamma$              & $\beta$                & $|Aut(C_i)|$          \\ \hline
\multirow{3}{*}{$C_{99}$} & \multirow{3}{*}{$W_{72,1}$} & $(1,1,1)$    & $(1,1,0)$    & $(0,1,1)$    & $(1,1,1)$    & $(1,1,0)$    & $(0,1,1)$    & $(1,0,0)$    & $(1,1,0)$    & $(1,1,0)$    & \multirow{3}{*}{$9$} & \multirow{3}{*}{$213$} & \multirow{3}{*}{$18$} \\
                          &                             & $r_{A_{10}}$ & $r_{A_{11}}$ & $r_{A_{12}}$ & $r_{A_{13}}$ & $r_{A_{14}}$ & $r_{A_{15}}$ & $r_{A_{16}}$ & $r_{A_{17}}$ & $r_{A_{18}}$ &                       &                        &                       \\
                          &                             & $(1,1,1)$      & $(0,1,1)$      & $(0,1,1)$      & $(1,1,1)$      & $(0,0,1)$      & $(1,1,0)$      & $(0,0,1)$      & $(0,1,0)$      & $(1,0,1)$      &                       &                        &                       \\ \hline
                        
\end{tabular}
\end{minipage}}
\end{table}

\begin{table}[h!]
\resizebox{0.65\textwidth}{!}{\begin{minipage}{\textwidth}
\centering
\begin{tabular}{cccccccccccccc}
\hline
                          & Type                        & $r_{A_1}$    & $r_{A_2}$    & $r_{A_3}$    & $r_{A_4}$    & $r_{A_5}$    & $r_{A_6}$    & $r_{A_7}$    & $r_{A_8}$    & $r_{A_9}$    & $\gamma$              & $\beta$                & $|Aut(C_i)|$          \\ \hline
\multirow{3}{*}{$C_{100}$} & \multirow{3}{*}{$W_{72,1}$} & $(0,1,1)$    & $(0,1,1)$    & $(1,0,1)$    & $(0,0,0)$    & $(0,0,0)$    & $(1,1,0)$    & $(1,0,0)$    & $(0,1,0)$    & $(0,0,1)$    & \multirow{3}{*}{$9$} & \multirow{3}{*}{$246$} & \multirow{3}{*}{$18$} \\
                          &                             & $r_{A_{10}}$ & $r_{A_{11}}$ & $r_{A_{12}}$ & $r_{A_{13}}$ & $r_{A_{14}}$ & $r_{A_{15}}$ & $r_{A_{16}}$ & $r_{A_{17}}$ & $r_{A_{18}}$ &                       &                        &                       \\
                          &                             & $(0,1,1)$      & $(1,0,1)$      & $(1,1,0)$      & $(1,0,0)$      & $(0,0,0)$      & $(0,1,1)$      & $(1,0,0)$      & $(1,1,0)$      & $(1,0,1)$      &                       &                        &                       \\ \hline
                        
\end{tabular}
\end{minipage}}
\end{table}

\begin{table}[h!]
\resizebox{0.65\textwidth}{!}{\begin{minipage}{\textwidth}
\centering
\begin{tabular}{cccccccccccccc}
\hline
                          & Type                        & $r_{A_1}$    & $r_{A_2}$    & $r_{A_3}$    & $r_{A_4}$    & $r_{A_5}$    & $r_{A_6}$    & $r_{A_7}$    & $r_{A_8}$    & $r_{A_9}$    & $\gamma$              & $\beta$                & $|Aut(C_i)|$          \\ \hline
\multirow{3}{*}{$C_{101}$} & \multirow{3}{*}{$W_{72,1}$} & $(0,1,0)$    & $(0,0,0)$    & $(0,0,0)$    & $(0,0,1)$    & $(0,1,0)$    & $(1,0,1)$    & $(0,1,0)$    & $(1,1,0)$    & $(0,1,0)$    & \multirow{3}{*}{$9$} & \multirow{3}{*}{$279$} & \multirow{3}{*}{$18$} \\
                          &                             & $r_{A_{10}}$ & $r_{A_{11}}$ & $r_{A_{12}}$ & $r_{A_{13}}$ & $r_{A_{14}}$ & $r_{A_{15}}$ & $r_{A_{16}}$ & $r_{A_{17}}$ & $r_{A_{18}}$ &                       &                        &                       \\
                          &                             & $(1,1,1)$      & $(1,0,0)$      & $(1,0,1)$      & $(0,1,0)$      & $(1,0,0)$      & $(0,1,0)$      & $(1,0,0)$      & $(0,1,0)$      & $(1,0,1)$      &                       &                        &                       \\ \hline
                        
\end{tabular}
\end{minipage}}
\end{table}

\begin{table}[h!]
\resizebox{0.65\textwidth}{!}{\begin{minipage}{\textwidth}
\centering
\begin{tabular}{cccccccccccccc}
\hline
                          & Type                        & $r_{A_1}$    & $r_{A_2}$    & $r_{A_3}$    & $r_{A_4}$    & $r_{A_5}$    & $r_{A_6}$    & $r_{A_7}$    & $r_{A_8}$    & $r_{A_9}$    & $\gamma$              & $\beta$                & $|Aut(C_i)|$          \\ \hline
\multirow{3}{*}{$C_{102}$} & \multirow{3}{*}{$W_{72,1}$} & $(1,0,1)$    & $(1,0,1)$    & $(1,1,0)$    & $(1,0,1)$    & $(1,1,0)$    & $(0,0,0)$    & $(1,0,1)$    & $(0,1,1)$    & $(0,1,1)$    & \multirow{3}{*}{$9$} & \multirow{3}{*}{$288$} & \multirow{3}{*}{$18$} \\
                          &                             & $r_{A_{10}}$ & $r_{A_{11}}$ & $r_{A_{12}}$ & $r_{A_{13}}$ & $r_{A_{14}}$ & $r_{A_{15}}$ & $r_{A_{16}}$ & $r_{A_{17}}$ & $r_{A_{18}}$ &                       &                        &                       \\
                          &                             & $(1,0,0)$      & $(0,0,1)$      & $(1,1,1)$      & $(0,1,0)$      & $(1,0,0)$      & $(1,0,0)$      & $(1,0,0)$      & $(0,0,1)$      & $(0,0,1)$      &                       &                        &                       \\ \hline
                        
\end{tabular}
\end{minipage}}
\end{table}

\newpage

\begin{table}[h!]
\resizebox{0.65\textwidth}{!}{\begin{minipage}{\textwidth}
\centering
\begin{tabular}{cccccccccccccc}
\hline
                          & Type                        & $r_{A_1}$    & $r_{A_2}$    & $r_{A_3}$    & $r_{A_4}$    & $r_{A_5}$    & $r_{A_6}$    & $r_{A_7}$    & $r_{A_8}$    & $r_{A_9}$    & $\gamma$              & $\beta$                & $|Aut(C_i)|$          \\ \hline
\multirow{3}{*}{$C_{103}$} & \multirow{3}{*}{$W_{72,1}$} & $(1,1,1)$    & $(1,0,0)$    & $(0,1,0)$    & $(1,1,1)$    & $(1,0,0)$    & $(0,0,1)$    & $(0,1,0)$    & $(1,1,1)$    & $(1,0,0)$    & \multirow{3}{*}{$9$} & \multirow{3}{*}{$315$} & \multirow{3}{*}{$18$} \\
                          &                             & $r_{A_{10}}$ & $r_{A_{11}}$ & $r_{A_{12}}$ & $r_{A_{13}}$ & $r_{A_{14}}$ & $r_{A_{15}}$ & $r_{A_{16}}$ & $r_{A_{17}}$ & $r_{A_{18}}$ &                       &                        &                       \\
                          &                             & $(1,1,1)$      & $(1,1,1)$      & $(0,1,0)$      & $(0,0,1)$      & $(1,1,1)$      & $(0,1,0)$      & $(0,0,1)$      & $(1,1,1)$      & $(0,0,1)$      &                       &                        &                       \\ \hline
                        
\end{tabular}
\end{minipage}}
\end{table}

\begin{table}[h!]
\resizebox{0.65\textwidth}{!}{\begin{minipage}{\textwidth}
\centering
\begin{tabular}{cccccccccccccc}
\hline
                          & Type                        & $r_{A_1}$    & $r_{A_2}$    & $r_{A_3}$    & $r_{A_4}$    & $r_{A_5}$    & $r_{A_6}$    & $r_{A_7}$    & $r_{A_8}$    & $r_{A_9}$    & $\gamma$              & $\beta$                & $|Aut(C_i)|$          \\ \hline
\multirow{3}{*}{$C_{104}$} & \multirow{3}{*}{$W_{72,1}$} & $(0,1,1)$    & $(0,1,0)$    & $(0,0,0)$    & $(0,0,1)$    & $(0,1,1)$    & $(1,0,0)$    & $(1,1,1)$    & $(1,1,1)$    & $(1,1,1)$    & \multirow{3}{*}{$9$} & \multirow{3}{*}{$324$} & \multirow{3}{*}{$18$} \\
                          &                             & $r_{A_{10}}$ & $r_{A_{11}}$ & $r_{A_{12}}$ & $r_{A_{13}}$ & $r_{A_{14}}$ & $r_{A_{15}}$ & $r_{A_{16}}$ & $r_{A_{17}}$ & $r_{A_{18}}$ &                       &                        &                       \\
                          &                             & $(0,1,0)$      & $(0,0,0)$      & $(1,0,0)$      & $(0,1,1)$      & $(0,1,0)$      & $(0,0,0)$      & $(1,0,1)$      & $(1,0,1)$      & $(0,1,1)$      &                       &                        &                       \\ \hline
                        
\end{tabular}
\end{minipage}}
\end{table}

\begin{table}[h!]
\resizebox{0.65\textwidth}{!}{\begin{minipage}{\textwidth}
\centering
\begin{tabular}{cccccccccccccc}
\hline
                          & Type                        & $r_{A_1}$    & $r_{A_2}$    & $r_{A_3}$    & $r_{A_4}$    & $r_{A_5}$    & $r_{A_6}$    & $r_{A_7}$    & $r_{A_8}$    & $r_{A_9}$    & $\gamma$              & $\beta$                & $|Aut(C_i)|$          \\ \hline
\multirow{3}{*}{$C_{105}$} & \multirow{3}{*}{$W_{72,1}$} & $(0,1,0)$    & $(0,1,0)$    & $(1,1,0)$    & $(0,0,0)$    & $(0,1,0)$    & $(0,0,0)$    & $(0,1,0)$    & $(0,1,1)$    & $(1,0,1)$    & \multirow{3}{*}{$9$} & \multirow{3}{*}{$354$} & \multirow{3}{*}{$18$} \\
                          &                             & $r_{A_{10}}$ & $r_{A_{11}}$ & $r_{A_{12}}$ & $r_{A_{13}}$ & $r_{A_{14}}$ & $r_{A_{15}}$ & $r_{A_{16}}$ & $r_{A_{17}}$ & $r_{A_{18}}$ &                       &                        &                       \\
                          &                             & $(1,1,1)$      & $(1,0,1)$      & $(0,0,0)$      & $(1,1,0)$      & $(1,1,1)$      & $(1,1,0)$      & $(1,0,1)$      & $(1,1,0)$      & $(1,0,1)$      &                       &                        &                       \\ \hline
                        
\end{tabular}
\end{minipage}}
\end{table}

\begin{table}[h!]
\resizebox{0.65\textwidth}{!}{\begin{minipage}{\textwidth}
\centering
\begin{tabular}{cccccccccccccc}
\hline
                          & Type                        & $r_{A_1}$    & $r_{A_2}$    & $r_{A_3}$    & $r_{A_4}$    & $r_{A_5}$    & $r_{A_6}$    & $r_{A_7}$    & $r_{A_8}$    & $r_{A_9}$    & $\gamma$              & $\beta$                & $|Aut(C_i)|$          \\ \hline
\multirow{3}{*}{$C_{106}$} & \multirow{3}{*}{$W_{72,1}$} & $(0,0,0)$    & $(1,1,0)$    & $(1,1,0)$    & $(0,0,1)$    & $(1,0,1)$    & $(1,0,0)$    & $(1,0,1)$    & $(1,1,1)$    & $(0,1,0)$    & \multirow{3}{*}{$9$} & \multirow{3}{*}{$357$} & \multirow{3}{*}{$18$} \\
                          &                             & $r_{A_{10}}$ & $r_{A_{11}}$ & $r_{A_{12}}$ & $r_{A_{13}}$ & $r_{A_{14}}$ & $r_{A_{15}}$ & $r_{A_{16}}$ & $r_{A_{17}}$ & $r_{A_{18}}$ &                       &                        &                       \\
                          &                             & $(0,1,0)$      & $(0,1,1)$      & $(1,1,1)$      & $(0,0,0)$      & $(1,1,1)$      & $(0,1,1)$      & $(0,0,1)$      & $(1,1,1)$      & $(1,1,1)$      &                       &                        &                       \\ \hline
                        
\end{tabular}
\end{minipage}}
\end{table}

\begin{table}[h!]
\resizebox{0.65\textwidth}{!}{\begin{minipage}{\textwidth}
\centering
\begin{tabular}{cccccccccccccc}
\hline
                          & Type                        & $r_{A_1}$    & $r_{A_2}$    & $r_{A_3}$    & $r_{A_4}$    & $r_{A_5}$    & $r_{A_6}$    & $r_{A_7}$    & $r_{A_8}$    & $r_{A_9}$    & $\gamma$              & $\beta$                & $|Aut(C_i)|$          \\ \hline
\multirow{3}{*}{$C_{107}$} & \multirow{3}{*}{$W_{72,1}$} & $(0,0,0)$    & $(1,1,0)$    & $(0,1,1)$    & $(1,1,1)$    & $(1,0,1)$    & $(1,1,0)$    & $(1,1,1)$    & $(0,0,0)$    & $(1,1,0)$    & \multirow{3}{*}{$27$} & \multirow{3}{*}{$354$} & \multirow{3}{*}{$18$} \\
                          &                             & $r_{A_{10}}$ & $r_{A_{11}}$ & $r_{A_{12}}$ & $r_{A_{13}}$ & $r_{A_{14}}$ & $r_{A_{15}}$ & $r_{A_{16}}$ & $r_{A_{17}}$ & $r_{A_{18}}$ &                       &                        &                       \\
                          &                             & $(1,1,1)$      & $(0,0,1)$      & $(0,0,1)$      & $(0,0,1)$      & $(0,1,0)$      & $(1,0,0)$      & $(0,0,0)$      & $(0,1,1)$      & $(0,0,0)$      &                       &                        &                       \\ \hline
                        
\end{tabular}
\end{minipage}}
\end{table}

\begin{table}[h!]
\resizebox{0.65\textwidth}{!}{\begin{minipage}{\textwidth}
\centering
\begin{tabular}{cccccccccccccc}
\hline
                          & Type                        & $r_{A_1}$    & $r_{A_2}$    & $r_{A_3}$    & $r_{A_4}$    & $r_{A_5}$    & $r_{A_6}$    & $r_{A_7}$    & $r_{A_8}$    & $r_{A_9}$    & $\gamma$              & $\beta$                & $|Aut(C_i)|$          \\ \hline
\multirow{3}{*}{$C_{108}$} & \multirow{3}{*}{$W_{72,1}$} & $(0,1,1)$    & $(0,1,1)$    & $(1,0,1)$    & $(0,1,1)$    & $(0,0,1)$    & $(0,1,1)$    & $(1,1,1)$    & $(1,1,0)$    & $(1,0,0)$    & \multirow{3}{*}{$27$} & \multirow{3}{*}{$405$} & \multirow{3}{*}{$18$} \\
                          &                             & $r_{A_{10}}$ & $r_{A_{11}}$ & $r_{A_{12}}$ & $r_{A_{13}}$ & $r_{A_{14}}$ & $r_{A_{15}}$ & $r_{A_{16}}$ & $r_{A_{17}}$ & $r_{A_{18}}$ &                       &                        &                       \\
                          &                             & $(1,0,0)$      & $(0,1,1)$      & $(0,1,0)$      & $(0,0,1)$      & $(0,0,0)$      & $(1,0,0)$      & $(1,0,0)$      & $(0,1,0)$      & $(1,0,0)$      &                       &                        &                       \\ \hline
                        
\end{tabular}
\end{minipage}}
\end{table}

\begin{table}[h!]
\resizebox{0.65\textwidth}{!}{\begin{minipage}{\textwidth}
\centering
\begin{tabular}{cccccccccccccc}
\hline
                          & Type                        & $r_{A_1}$    & $r_{A_2}$    & $r_{A_3}$    & $r_{A_4}$    & $r_{A_5}$    & $r_{A_6}$    & $r_{A_7}$    & $r_{A_8}$    & $r_{A_9}$    & $\gamma$              & $\beta$                & $|Aut(C_i)|$          \\ \hline
\multirow{3}{*}{$C_{109}$} & \multirow{3}{*}{$W_{72,1}$} & $(0,0,1)$    & $(0,1,1)$    & $(0,1,1)$    & $(0,1,0)$    & $(0,1,0)$    & $(1,1,0)$    & $(0,0,0)$    & $(0,0,0)$    & $(1,1,0)$    & \multirow{3}{*}{$27$} & \multirow{3}{*}{$444$} & \multirow{3}{*}{$18$} \\
                          &                             & $r_{A_{10}}$ & $r_{A_{11}}$ & $r_{A_{12}}$ & $r_{A_{13}}$ & $r_{A_{14}}$ & $r_{A_{15}}$ & $r_{A_{16}}$ & $r_{A_{17}}$ & $r_{A_{18}}$ &                       &                        &                       \\
                          &                             & $(0,0,1)$      & $(0,0,1)$      & $(0,0,0)$      & $(1,1,0)$      & $(1,0,1)$      & $(1,0,0)$      & $(0,0,1)$      & $(1,0,0)$      & $(0,0,1)$      &                       &                        &                       \\ \hline
                        
\end{tabular}
\end{minipage}}
\end{table}

\item[2.] Generator matrices $\mathcal{G}_2, \mathcal{G}_2'$ and $\mathcal{G}_2''$

In the generator matrix $\mathcal{G}_2,$ the matrix $\tau_2(v_2)$ is fully defined by the first row, for this reason, we only list the first row of the matrices $A$ and $B$ which we label as $r_{A}$ and $r_{B}$ respectively.

\begin{table}[h!]\label{Case 2}
\caption{New Type I $[72,36,12]$ Codes from $\mathcal{G}_2$ and $R=\mathbb{F}_2$}
\resizebox{0.65\textwidth}{!}{\begin{minipage}{\textwidth}
\centering
\begin{tabular}{ccccccc}
\hline
      & Type       & $r_A$                                   & $r_B$                   & $\gamma$ & $\beta$ & $|Aut(C_i)|$ \\ \hline
$C_{110}$ & $W_{72,1}$ & $(0, 0, 1, 0, 0, 1, 0, 1, 0, 0, 1, 0, 1, 0, 1, 0, 0, 0)$ & $( 1, 1, 0, 1, 0, 0, 0, 0, 0, 0, 0, 1, 0, 0, 1, 1, 1, 0 )$ & $0$      & $270$   & $36$         \\ \hline
$C_{111}$ & $W_{72,1}$ & $(0,1,1,1,1,1,0,0,1,0,1,1,0,1,1,0,1,1)$ & $(1,0,1,0,0,0,0,0,1,1,0,1,0,1,1,0,1,1)$ & $18$      & $216$   & $36$         \\ \hline
$C_{112}$ & $W_{72,1}$ & $(0,1,0,1,0,0,1,1,1,0,0,0,0,0,1,1,1,0)$ & $(1,0,0,1,1,0,0,0,0,1,1,1,0,0,1,0,1,1)$ & $18$      & $249$   & $36$         \\ \hline
$C_{113}$ & $W_{72,1}$ & $(0, 0, 0, 0, 0, 1, 1, 0, 1, 1, 1, 1, 0, 1, 0, 0, 1, 0)$ & $(0, 1, 0, 1, 0, 1, 1, 1, 1, 0, 1, 1, 1, 1, 1, 1, 1, 0)$ & $18$      & $309$   & $36$         \\ \hline
$C_{114}$ & $W_{72,1}$ & $(1, 1, 1, 0, 1, 0, 1, 0, 1, 0, 0, 0, 0, 1, 1, 1, 0, 1)$ & $(1, 1, 1, 0, 1, 0, 0, 0, 1, 1, 0, 0, 0, 0, 0, 0, 0, 1)$ & $18$      & $315$   & $36$         \\ \hline
$C_{115}$ & $W_{72,1}$ & $(1, 0, 0, 0, 0, 0, 0, 1, 0, 1, 0, 0, 0, 1, 1, 1, 0, 1)$ & $(1, 0, 0, 1, 1, 0, 1, 1, 1, 1, 1, 1, 0, 0, 1, 0, 0, 0)$ & $18$      & $327$   & $36$         \\ \hline
$C_{116}$ & $W_{72,1}$ & $(0, 0, 0, 1,1, 0, 1, 0, 1, 1, 0, 1, 1, 1, 1, 0, 1, 0)$ & $(1, 1, 1, 1, 0, 0, 0, 1, 0, 1, 0, 0, 0, 0, 0, 0, 0, 1)$ & $18$      & $354$   & $36$         \\ \hline
$C_{117}$ & $W_{72,1}$ & $(0,1,0,1,0,0,1,1,1,0,0,0,0,0,1,1,1,0)$ & $(1,0,1,0,1,0,0,0,0,1,1,1,0,0,1,0,1,1)$ & $18$      & $435$   & $36$         \\ \hline
$C_{118}$ & $W_{72,1}$ & $(1,1,1,1,0,0,1,1,0,0,0,1,1,1,1,0,1,0)$ & $(1,1,0,1,1,0,0,1,1,0,1,1,1,0,0,0,1,0)$ & $18$      & $468$   & $36$         \\ \hline
$C_{119}$ & $W_{72,1}$ & $(1, 0, 1, 0, 0, 0, 1, 1, 1, 1, 0, 1, 1, 1, 1, 0, 1, 0)$ & $(1, 1, 0, 0, 1, 0, 1, 0, 0, 0, 0, 0, 0, 1, 0, 0, 1, 0)$ & $36$      & $420$   & $36$         \\ \hline
$C_{120}$ & $W_{72,1}$ & $(1,1,0,1,0,0,1,1,0,0,1,1,1,1,1,0,1,0)$ & $(1,1,0,1,1,1,0,1,1,0,1,0,1,0,0,0,1,0)$ & $36$      & $567$   & $36$         \\ \hline
$C_{121}$ & $W_{72,1}$ & $(1, 0, 0, 0, 1, 1, 0, 1, 1, 1, 1, 0, 1, 0, 1, 0, 0, 1)$ & $(1, 0, 1, 1, 1, 0, 0, 0, 1, 0, 1, 1, 0, 0, 0, 0, 0, 0)$ & $36$      & $615$   & $36$         \\ \hline
$C_{122}$ & $W_{72,1}$ & $(1, 0, 0, 1, 0, 1, 1, 0, 1, 0, 1, 1, 0, 1, 1, 1, 1, 1)$ & $(0, 0, 0, 0, 0, 0, 0, 0, 1, 0, 0, 0, 0, 1, 1, 1, 1, 0)$ & $54$      & $642$   & $36$         \\ \hline
\end{tabular}
\end{minipage}}
\end{table}
\newpage
In the generator matrix $\mathcal{G}_2',$ the matrix $\tau_2(v_2)$ is fully defined by the $2 \times 2$ matrices in the first row- some of them are circulant and some of them are persymmetric. For this reason, we only list the first row of the matrices $A_1,A_2,A_3,\dots,A_{18}$ which we label as $r_{A_1},r_{A_2},r_{A_3},\dots,r_{A_{18}}$ respectively. If the matrix $A_i$ is circulant, we only list the first row of such matrix and if the matrix $A_i$ is persymmetric, we only list the three variables that correspond to such matrix.

\begin{table}[h!]
\caption{New Type~I $[72,36,12]$ Codes from $\mathcal{G}_2'$ and $R=\mathbb{F}_2$}
\resizebox{0.65\textwidth}{!}{\begin{minipage}{\textwidth}
\centering
\begin{tabular}{cccccccccccccc}
\hline
                          & Type                        & $r_{A_1}$    & $r_{A_2}$    & $r_{A_3}$    & $r_{A_4}$    & $r_{A_5}$    & $r_{A_6}$    & $r_{A_7}$    & $r_{A_8}$    & $r_{A_9}$    & $\gamma$              & $\beta$                & $|Aut(C_i)|$          \\ \hline
\multirow{3}{*}{$C_{123}$} & \multirow{3}{*}{$W_{72,1}$} & $(1,1,0)$    & $(1,0,0)$    & $(1,0,1)$    & $(1,0,1)$    & $(1,1,1)$    & $(0,0,1)$    & $(0,1,0)$    & $(1,0,0)$    & $(1,1,0)$    & \multirow{3}{*}{$9$} & \multirow{3}{*}{$273$} & \multirow{3}{*}{$18$} \\
                          &                             & $r_{A_{10}}$ & $r_{A_{11}}$ & $r_{A_{12}}$ & $r_{A_{13}}$ & $r_{A_{14}}$ & $r_{A_{15}}$ & $r_{A_{16}}$ & $r_{A_{17}}$ & $r_{A_{18}}$ &                       &                        &                       \\
                          &                             & $(0,0)$      & $(1,0)$      & $(0,0)$      & $(0,0)$      & $(1,1)$      & $(1,0)$      & $(0,0)$      & $(1,1)$      & $(0,0)$      &                       &                        &                       \\ \hline
                        
\end{tabular}
\end{minipage}}
\end{table}

\begin{table}[h!]
\resizebox{0.65\textwidth}{!}{\begin{minipage}{\textwidth}
\centering
\begin{tabular}{cccccccccccccc}
\hline
                          & Type                        & $r_{A_1}$    & $r_{A_2}$    & $r_{A_3}$    & $r_{A_4}$    & $r_{A_5}$    & $r_{A_6}$    & $r_{A_7}$    & $r_{A_8}$    & $r_{A_9}$    & $\gamma$              & $\beta$                & $|Aut(C_i)|$          \\ \hline
\multirow{3}{*}{$C_{124}$} & \multirow{3}{*}{$W_{72,1}$} & $(1,1,0)$    & $(1,1,0)$    & $(1,0,0)$    & $(0,0,1)$    & $(0,1,0)$    & $(1,1,1)$    & $(0,0,1)$    & $(0,0,1)$    & $(0,0,0)$    & \multirow{3}{*}{$9$} & \multirow{3}{*}{$297$} & \multirow{3}{*}{$18$} \\
                          &                             & $r_{A_{10}}$ & $r_{A_{11}}$ & $r_{A_{12}}$ & $r_{A_{13}}$ & $r_{A_{14}}$ & $r_{A_{15}}$ & $r_{A_{16}}$ & $r_{A_{17}}$ & $r_{A_{18}}$ &                       &                        &                       \\
                          &                             & $(1,1)$      & $(1,1)$      & $(0,0)$      & $(0,1)$      & $(0,1)$      & $(0,0)$      & $(0,1)$      & $(0,1)$      & $(0,1)$      &                       &                        &                       \\ \hline
                        
\end{tabular}
\end{minipage}}
\end{table}

\begin{table}[h!]
\resizebox{0.65\textwidth}{!}{\begin{minipage}{\textwidth}
\centering
\begin{tabular}{cccccccccccccc}
\hline
                          & Type                        & $r_{A_1}$    & $r_{A_2}$    & $r_{A_3}$    & $r_{A_4}$    & $r_{A_5}$    & $r_{A_6}$    & $r_{A_7}$    & $r_{A_8}$    & $r_{A_9}$    & $\gamma$              & $\beta$                & $|Aut(C_i)|$          \\ \hline
\multirow{3}{*}{$C_{125}$} & \multirow{3}{*}{$W_{72,1}$} & $(1,1,0)$    & $(0,1,1)$    & $(0,1,0)$    & $(0,0,1)$    & $(0,0,0)$    & $(1,1,0)$    & $(0,1,0)$    & $(0,0,0)$    & $(0,1,0)$    & \multirow{3}{*}{$9$} & \multirow{3}{*}{$306$} & \multirow{3}{*}{$18$} \\
                          &                             & $r_{A_{10}}$ & $r_{A_{11}}$ & $r_{A_{12}}$ & $r_{A_{13}}$ & $r_{A_{14}}$ & $r_{A_{15}}$ & $r_{A_{16}}$ & $r_{A_{17}}$ & $r_{A_{18}}$ &                       &                        &                       \\
                          &                             & $(1,0)$      & $(0,0)$      & $(0,1)$      & $(0,0)$      & $(1,1)$      & $(1,1)$      & $(0,1)$      & $(0,0)$      & $(1,1)$      &                       &                        &                       \\ \hline
                        
\end{tabular}
\end{minipage}}
\end{table}

\begin{table}[h!]
\resizebox{0.65\textwidth}{!}{\begin{minipage}{\textwidth}
\centering
\begin{tabular}{cccccccccccccc}
\hline
                          & Type                        & $r_{A_1}$    & $r_{A_2}$    & $r_{A_3}$    & $r_{A_4}$    & $r_{A_5}$    & $r_{A_6}$    & $r_{A_7}$    & $r_{A_8}$    & $r_{A_9}$    & $\gamma$              & $\beta$                & $|Aut(C_i)|$          \\ \hline
\multirow{3}{*}{$C_{126}$} & \multirow{3}{*}{$W_{72,1}$} & $(0,1,0)$    & $(1,1,0)$    & $(1,1,1)$    & $(0,0,1)$    & $(1,0,1)$    & $(1,0,0)$    & $(0,0,1)$    & $(0,1,0)$    & $(1,0,0)$    & \multirow{3}{*}{$9$} & \multirow{3}{*}{$309$} & \multirow{3}{*}{$18$} \\
                          &                             & $r_{A_{10}}$ & $r_{A_{11}}$ & $r_{A_{12}}$ & $r_{A_{13}}$ & $r_{A_{14}}$ & $r_{A_{15}}$ & $r_{A_{16}}$ & $r_{A_{17}}$ & $r_{A_{18}}$ &                       &                        &                       \\
                          &                             & $(1,0)$      & $(0,0)$      & $(0,0)$      & $(1,0)$      & $(0,1)$      & $(1,0)$      & $(1,0)$      & $(1,1)$      & $(1,0)$      &                       &                        &                       \\ \hline
                        
\end{tabular}
\end{minipage}}
\end{table}

In the generator matrix $\mathcal{G}_2'',$ the matrix $\tau_2(v_2)$ is fully defined by the $2 \times 2$ matrices in the first row that are all persymmetric. For this reason, we only list the first row of the matrices $A_1,A_2,A_3,\dots,A_{18}$ which we label as $r_{A_1},r_{A_2},r_{A_3},\dots,r_{A_{18}}$ respectively. We only list the three variables that correspond to such matrix.

\begin{table}[h!]
\caption{New Type~I $[72,36,12]$ Codes from $\mathcal{G}_2''$ and $R=\mathbb{F}_2$}
\resizebox{0.65\textwidth}{!}{\begin{minipage}{\textwidth}
\centering
\begin{tabular}{cccccccccccccc}
\hline
                          & Type                        & $r_{A_1}$    & $r_{A_2}$    & $r_{A_3}$    & $r_{A_4}$    & $r_{A_5}$    & $r_{A_6}$    & $r_{A_7}$    & $r_{A_8}$    & $r_{A_9}$    & $\gamma$              & $\beta$                & $|Aut(C_i)|$          \\ \hline
\multirow{3}{*}{$C_{127}$} & \multirow{3}{*}{$W_{72,1}$} & $(0,0,1)$    & $(1,0,1)$    & $(1,0,0)$    & $(1,1,1)$    & $(1,0,0)$    & $(1,1,0)$    & $(1,1,0)$    & $(1,0,1)$    & $(1,1,0)$    & \multirow{3}{*}{$0$} & \multirow{3}{*}{$465$} & \multirow{3}{*}{$36$} \\
                          &                             & $r_{A_{10}}$ & $r_{A_{11}}$ & $r_{A_{12}}$ & $r_{A_{13}}$ & $r_{A_{14}}$ & $r_{A_{15}}$ & $r_{A_{16}}$ & $r_{A_{17}}$ & $r_{A_{18}}$ &                       &                        &                       \\
                          &                             & $(1,0,0)$      & $(0,0,0)$      & $(1,0,1)$      & $(0,1,1)$      & $(1,0,0)$      & $(1,0,0)$      & $(1,0,0)$      & $(1,1,1)$      & $(0,1,0)$      &                       &                        &                       \\ \hline
                        
\end{tabular}
\end{minipage}}
\end{table}

\begin{table}[h!]
\resizebox{0.65\textwidth}{!}{\begin{minipage}{\textwidth}
\centering
\begin{tabular}{cccccccccccccc}
\hline
                          & Type                        & $r_{A_1}$    & $r_{A_2}$    & $r_{A_3}$    & $r_{A_4}$    & $r_{A_5}$    & $r_{A_6}$    & $r_{A_7}$    & $r_{A_8}$    & $r_{A_9}$    & $\gamma$              & $\beta$                & $|Aut(C_i)|$          \\ \hline
\multirow{3}{*}{$C_{128}$} & \multirow{3}{*}{$W_{72,1}$} & $(1,1,0)$    & $(0,0,1)$    & $(0,0,0)$    & $(1,0,1)$    & $(0,0,0)$    & $(0,1,1)$    & $(0,0,0)$    & $(1,1,0)$    & $(1,0,0)$    & \multirow{3}{*}{$9$} & \multirow{3}{*}{$342$} & \multirow{3}{*}{$18$} \\
                          &                             & $r_{A_{10}}$ & $r_{A_{11}}$ & $r_{A_{12}}$ & $r_{A_{13}}$ & $r_{A_{14}}$ & $r_{A_{15}}$ & $r_{A_{16}}$ & $r_{A_{17}}$ & $r_{A_{18}}$ &                       &                        &                       \\
                          &                             & $(0,1,1)$      & $(1,1,1)$      & $(1,0,0)$      & $(0,1,0)$      & $(0,0,0)$      & $(1,1,0)$      & $(0,1,0)$      & $(0,1,0)$      & $(1,0,0)$      &                       &                        &                       \\ \hline
                        
\end{tabular}
\end{minipage}}
\end{table}

\begin{table}[h!]
\resizebox{0.65\textwidth}{!}{\begin{minipage}{\textwidth}
\centering
\begin{tabular}{cccccccccccccc}
\hline
                          & Type                        & $r_{A_1}$    & $r_{A_2}$    & $r_{A_3}$    & $r_{A_4}$    & $r_{A_5}$    & $r_{A_6}$    & $r_{A_7}$    & $r_{A_8}$    & $r_{A_9}$    & $\gamma$              & $\beta$                & $|Aut(C_i)|$          \\ \hline
\multirow{3}{*}{$C_{129}$} & \multirow{3}{*}{$W_{72,1}$} & $(0,0,0)$    & $(1,0,0)$    & $(0,0,0)$    & $(0,1,1)$    & $(0,0,0)$    & $(1,1,1)$    & $(0,1,1)$    & $(1,0,0)$    & $(0,1,1)$    & \multirow{3}{*}{$9$} & \multirow{3}{*}{$345$} & \multirow{3}{*}{$18$} \\
                          &                             & $r_{A_{10}}$ & $r_{A_{11}}$ & $r_{A_{12}}$ & $r_{A_{13}}$ & $r_{A_{14}}$ & $r_{A_{15}}$ & $r_{A_{16}}$ & $r_{A_{17}}$ & $r_{A_{18}}$ &                       &                        &                       \\
                          &                             & $(0,1,1)$      & $(0,1,0)$      & $(1,1,0)$      & $(1,1,1)$      & $(1,1,0)$      & $(1,1,0)$      & $(1,1,1)$      & $(0,0,1)$      & $(1,1,0)$      &                       &                        &                       \\ \hline
                        
\end{tabular}
\end{minipage}}
\end{table}

\begin{table}[h!]
\resizebox{0.65\textwidth}{!}{\begin{minipage}{\textwidth}
\centering
\begin{tabular}{cccccccccccccc}
\hline
                          & Type                        & $r_{A_1}$    & $r_{A_2}$    & $r_{A_3}$    & $r_{A_4}$    & $r_{A_5}$    & $r_{A_6}$    & $r_{A_7}$    & $r_{A_8}$    & $r_{A_9}$    & $\gamma$              & $\beta$                & $|Aut(C_i)|$          \\ \hline
\multirow{3}{*}{$C_{130}$} & \multirow{3}{*}{$W_{72,1}$} & $(0,0,0)$    & $(1,0,0)$    & $(1,0,1)$    & $(1,1,1)$    & $(1,0,1)$    & $(0,1,1)$    & $(1,0,0)$    & $(0,1,0)$    & $(1,1,0)$    & \multirow{3}{*}{$9$} & \multirow{3}{*}{$375$} & \multirow{3}{*}{$18$} \\
                          &                             & $r_{A_{10}}$ & $r_{A_{11}}$ & $r_{A_{12}}$ & $r_{A_{13}}$ & $r_{A_{14}}$ & $r_{A_{15}}$ & $r_{A_{16}}$ & $r_{A_{17}}$ & $r_{A_{18}}$ &                       &                        &                       \\
                          &                             & $(1,0,1)$      & $(1,1,0)$      & $(1,1,1)$      & $(0,0,1)$      & $(1,0,0)$      & $(1,1,0)$      & $(1,1,1)$      & $(0,0,1)$      & $(0,1,0)$      &                       &                        &                       \\ \hline
                        
\end{tabular}
\end{minipage}}
\end{table}

\item[3.] Generator matrix $\mathcal{G}_3$

In the generator matrix $\mathcal{G}_3,$ the matrix $\tau_2(v_3)$ is fully defined by the first row, for this reason, we only list the first row of the matrices $A,B,C,D,E$ and $F$ which we label as $r_{A},r_{B},r_{C},r_{D},r_{E}$ and $r_{F}$ respectively.

\begin{table}[h!]\label{C18}
\caption{New Type I $[72,36,12]$ Codes from $\mathcal{G}_3$ and $R=\mathbb{F}_2$}
\resizebox{0.6\textwidth}{!}{\begin{minipage}{\textwidth}
\centering
\begin{tabular}{ccccccccccc}
\hline
 & Type       & $r_A$       & $r_B$ & $r_C$     & $r_D$    & $r_E$       & $r_F$         & $\gamma$ & $\beta$ & $|Aut(C_i)|$ \\ \hline
$C_{131}$ & $W_{72,1}$ & $(1, 0, 0, 0, 1, 1)$ & $(0, 1, 1, 1, 0, 1)$ &$(1, 0, 1, 1, 1, 0)$ &$(1, 0, 0, 0, 1, 0)$ &$(0, 1, 1, 0, 0, 0)$ &$(1, 1, 1, 1, 1, 1)$  & $0$      & $135$   & $72$         \\ \hline
\end{tabular}
\end{minipage}}
\end{table}

\item[4.] Generator matrix $\mathcal{G}_4$

In the generator matrix $\mathcal{G}_4,$ the matrix $\tau_2(v_4)$ is fully defined by the first row, for this reason, we only list the first row of the matrices $A,B$ and $C$ which we label as $r_{A},r_{B}$ and $r_{C}$ respectively.

\begin{table}[h!]\label{Case C3XC6}
\caption{New Type I $[72,36,12]$ Codes from $\mathcal{G}_4$ and $R=\mathbb{F}_2$}
\resizebox{0.65\textwidth}{!}{\begin{minipage}{\textwidth}
\centering
\begin{tabular}{cccccccc}
\hline
         & Type       & $r_A$                                  & $r_B$                                   & $r_C$                                   & $\gamma$ & $\beta$ & $|Aut(C_i)|$ \\ \hline
$C_{132}$ & $W_{72,1}$ & $(1, 1, 1, 1, 0, 1, 1, 1, 0, 0, 0, 0)$ & $( 0, 1, 0, 1, 0, 0, 1, 0, 1, 1, 0, 1)$ &  $(1, 0, 0, 1, 1, 1, 1, 0, 1, 1, 0, 1)$ &  $0$      & $165$  & $72$         \\ \hline
\end{tabular}
\end{minipage}}
\end{table}

In the generator matrix $\mathcal{G}_4',$ the matrix $\tau_2(v_4)$ is fully defined by the $2 \times 2$ matrices in the first row- some of them are circulant and some of them are persymmetric. For this reason, we only list the first row of the matrices $A_1,A_2,A_3,\dots,A_{18}$ which we label as $r_{A_1},r_{A_2},r_{A_3},\dots,r_{A_{18}}$ respectively. If the matrix $A_i$ is circulant, we only list the first row of such matrix and if the matrix $A_i$ is persymmetric, we only list the three variables that correspond to such matrix.

\begin{table}[h!]
\caption{New Type~I $[72,36,12]$ Codes from $\mathcal{G}_4'$ and $R=\mathbb{F}_2$}
\resizebox{0.65\textwidth}{!}{\begin{minipage}{\textwidth}
\centering
\begin{tabular}{cccccccccccccc}
\hline
                          & Type                        & $r_{A_1}$    & $r_{A_2}$    & $r_{A_3}$    & $r_{A_4}$    & $r_{A_5}$    & $r_{A_6}$    & $r_{A_7}$    & $r_{A_8}$    & $r_{A_9}$    & $\gamma$              & $\beta$                & $|Aut(C_i)|$          \\ \hline
\multirow{3}{*}{$C_{133}$} & \multirow{3}{*}{$W_{72,1}$} & $(0,1,0)$    & $(1,0,1)$    & $(0,1,0)$    & $(0,0,0)$    & $(1,0,0)$    & $(1,0,0)$    & $(1,0,1)$    & $(1,1,0)$    & $(1,0,1)$    & \multirow{3}{*}{$18$} & \multirow{3}{*}{$297$} & \multirow{3}{*}{$36$} \\
                          &                             & $r_{A_{10}}$ & $r_{A_{11}}$ & $r_{A_{12}}$ & $r_{A_{13}}$ & $r_{A_{14}}$ & $r_{A_{15}}$ & $r_{A_{16}}$ & $r_{A_{17}}$ & $r_{A_{18}}$ &                       &                        &                       \\
                          &                             & $(1,0)$      & $(1,0)$      & $(0,0)$      & $(0,0)$      & $(1,0)$      & $(1,1)$      & $(0,1)$      & $(1,1)$      & $(0,0)$      &                       &                        &                       \\ \hline
                        
\end{tabular}
\end{minipage}}
\end{table}

\begin{table}[h!]
\resizebox{0.65\textwidth}{!}{\begin{minipage}{\textwidth}
\centering
\begin{tabular}{cccccccccccccc}
\hline
                          & Type                        & $r_{A_1}$    & $r_{A_2}$    & $r_{A_3}$    & $r_{A_4}$    & $r_{A_5}$    & $r_{A_6}$    & $r_{A_7}$    & $r_{A_8}$    & $r_{A_9}$    & $\gamma$              & $\beta$                & $|Aut(C_i)|$          \\ \hline
\multirow{3}{*}{$C_{134}$} & \multirow{3}{*}{$W_{72,1}$} & $(1,1,0)$    & $(0,0,0)$    & $(0,1,1)$    & $(0,0,1)$    & $(0,0,1)$    & $(0,1,0)$    & $(0,0,0)$    & $(0,0,1)$    & $(1,1,0)$    & \multirow{3}{*}{$36$} & \multirow{3}{*}{$378$} & \multirow{3}{*}{$36$} \\
                          &                             & $r_{A_{10}}$ & $r_{A_{11}}$ & $r_{A_{12}}$ & $r_{A_{13}}$ & $r_{A_{14}}$ & $r_{A_{15}}$ & $r_{A_{16}}$ & $r_{A_{17}}$ & $r_{A_{18}}$ &                       &                        &                       \\
                          &                             & $(0,1)$      & $(1,1)$      & $(0,0)$      & $(1,0)$      & $(1,0)$      & $(1,1)$      & $(0,1)$      & $(1,0)$      & $(1,1)$      &                       &                        &                       \\ \hline
                        
\end{tabular}
\end{minipage}}
\end{table}

\item[5.] Generator matrix $\mathcal{G}_5$

In the generator matrix $\mathcal{G}_5,$ the matrix $\tau_2(v_5)$ is fully defined by the first row, for this reason, we only list the first row of the matrices $A,B,C,D,E$ and $F$ which we label as $r_{A},r_{B},r_{C},r_{D},r_{E}$ and $r_{E}$ respectively.

\begin{table}[h!]\label{Case C6XC3}
\caption{New Type II $[72,36,12]$ Codes from $\mathcal{G}_5$ and $R=\mathbb{F}_2$}
\resizebox{0.65\textwidth}{!}{\begin{minipage}{\textwidth}
\centering
\begin{tabular}{cccccccccc}
\hline
               & $r_A$                         & $r_B$                   & $r_C$             & $r_D$                    & $r_E$             & $r_F$             & $\alpha$ & $|Aut(C_i)|$ \\ \hline
$C_{135}$  & $(0, 0, 0, 1, 0, 0)$ & $( 0, 0, 1, 0, 0, 0)$ &  $(1, 0, 0, 0, 0, 0)$ & $( 0, 0, 0, 1, 1, 0)$ & $(  1, 0, 0, 1, 0, 1)$ &  $(0, 0, 1, 1, 0, 1 )$      & $-1980$  & $432$         \\ \hline
\end{tabular}
\end{minipage}}
\end{table}

\end{enumerate}

\section{Conclusion}

In this paper, we defined generator matrices of the form $[I_{kn} \ | \ \tau_k(v)]$ - this idea was first introduced in \cite{Dougherty3}. Such generator matrices depend on the choice of the group $G$ and the form of the $k \times k$ matrices. We specifically considered groups of order 18 and some $2 \times 2$ matrices, that is, $k=2$ in our generator matrices. We then employed our generator matrices to search for binary $[72,36,12]$ self-dual codes. We were able to construct Type I binary $[72,36,12]$ self-dual codes with new weight enumerators in $W_{72,1}$:

\begin{equation*}
\begin{array}{l}
(\gamma =0,\ \  \beta =\{93,111,132,135,138,144,150,165,174,198,270,282,309,345,\\
\qquad \qquad \qquad 366,378, 411,444,453,465 \}), \\

(\gamma =9,\ \  \beta =\{192,210,213,225,228,246,255,258,261,270,273,279,282,288,\\
\qquad \qquad \qquad 291, 297,300,306,309,315,318,324,336,342,345,354,357,375,\\
\qquad \qquad \qquad 393 \}), \\

(\gamma =18,\  \beta =\{216,228,243,249,252,255, 267, 282, 291, 294, 297,303, 309,312,\\
\qquad \qquad \qquad 315,318, 321, 327,330,333,339,348,351,354,360, 363,366, 369,\\
\qquad \qquad \qquad 372, 381,384, 390,393,399,402,408,411, 414, 417, 423, 426,435,\\
\qquad \qquad \qquad 438,444, 450, 462, 468, 471, 474,480,486,489,498,507,516,525,\\
\qquad \qquad \qquad 540\}), \\

(\gamma =27,\ \beta =\{354,405,444\}), \\

(\gamma =36,\ \beta =378,393,399,402,420, 444,453, 462, 477, 489, 507, 516, 525, 534,\\
\qquad \qquad \qquad 567, 582,588, 600, 606, 615, 624, 663 \}) \\

(\gamma =54,\ \beta =\{642,651,657\}) \\
\end{array}%
\end{equation*}
and Type II binary $[72,36,12]$ self-dual codes with new weight enumerators:
\begin{equation*}
\begin{array}{l}
(\alpha =\{-1980 \}), \\

\end{array}%
\end{equation*}

A suggestion for future work is to consider generator matrices of the form $[I_{kn} \ | \ \tau_k(v)]$ for groups of orders different than 18 and for values of $k$ different than 2, to search for optimal binary self-dual codes of different lengths. Another suggestion is to consider generator matrices of the form $[I_{kn} \ | \ \tau_k(v)]$ over different alphabets, for example, rings, and explore the binary images of the codes under the Gray maps.

\end{document}